\documentclass[11pt]{article}

\usepackage[utf8]{inputenx}
\usepackage[english]{babel}
\usepackage{amsmath}
\usepackage{amssymb}
\usepackage{hyperref}
\usepackage{float}
\usepackage{stix2}
\usepackage{titlesec}
\usepackage{multirow}
\usepackage{latexsym}
\usepackage{caption}
\usepackage{latexsym}
\usepackage[T1]{fontenc}
\usepackage{dcolumn}
\usepackage{eurosym}
\usepackage{setspace}
\usepackage{color}
\usepackage[usenames,dvipsnames]{xcolor}
\usepackage{geometry}
\usepackage{multicol}
\usepackage{caption}
\usepackage{amsthm}
\usepackage{stmaryrd}
\usepackage{enumitem}
\usepackage{physics}
\usepackage{bbm}
\usepackage{scalerel}
\usepackage{mathtools}
\usepackage{authblk}
\usepackage[titletoc]{appendix}
\usepackage{tocloft}
\usepackage{abstract}

\titleformat{\section}{\it\LARGE}{\thesection}{1em}{}
\titleformat{\subsection}{\it\Large}{\thesubsection}{1em}{}
\titleformat{\subsubsection}{\it\large}{\thesubsubsection}{1em}{}

\renewcommand\contentsname{\textit{Contents}}

\makeatletter
\renewcommand*\l@section[2]{%
  \ifnum \c@tocdepth >\m@ne
    \addpenalty{-\@highpenalty}%
    \vskip 1.0em \@plus\p@
    \setlength\@tempdima{1.5em}%
    \begingroup
      \parindent \z@ \rightskip \@pnumwidth
      \parfillskip -\@pnumwidth
      \leavevmode \it\large
      \advance\leftskip\@tempdima
      \hskip -\leftskip
      #1\nobreak\hfil \nobreak\hb@xt@\@pnumwidth{\hss\normalfont #2}\par
      \penalty\@highpenalty
    \endgroup
  \fi}
\makeatother

\usepackage[style=authoryear-comp,citestyle=authoryear-comp,backend=bibtex]{biblatex}
\DeclareNameAlias{sortname}{last-first}
\AtBeginDocument{\nocite{*}}
\DeclareFieldFormat[article,inbook,book,incollection,inproceedings,patent,thesis,unpublished,online,misc]{title}{``\textit{#1}''}

\makeatletter
\DeclareRobustCommand{\cev}[1]{%
  \mathpalette\do@cev{#1}%
}
\newcommand{\do@cev}[2]{%
  \fix@cev{#1}{+}%
  \reflectbox{$\m@th#1\vec{\reflectbox{$\fix@cev{#1}{-}\m@th#1#2\fix@cev{#1}{+}$}}$}%
  \fix@cev{#1}{-}%
}
\newcommand{\fix@cev}[2]{%
  \ifx#1\displaystyle
    \mkern#23mu
  \else
    \ifx#1\textstyle
      \mkern#23mu
    \else
      \ifx#1\scriptstyle
        \mkern#22mu
      \else
        \mkern#22mu
      \fi
    \fi
  \fi
}

\makeatother

\newcommand{\qf}[1]{\pmb{#1}}
\newcommand{\qs}[1]{\hat{#1}}

\newcommand{\te}[1]{\text{#1}}
\newcommand{\cfv}{\phi_0}
\newcommand{\cfm}{\mu_0}
\newcommand{\csm}{\rho}
\newcommand{\csv}{b}
\newcommand{\cmg}{m_{\te{G}}}
\newcommand{\ok}{\omega_{\te{G}}}
\newcommand{\ml}{m_{\te{KG}}}
\newcommand{\olam}{\omega_{\te{KG}}}
\newcommand{\qcfv}{\qf{\phi}_0}
\newcommand{\qcfm}{\qf{\mu}_0}

\newcommand{\Gv}{a}
\newcommand{\Gm}{p_a}
\newcommand{\KGm}{p_\phi}
\newcommand{\KGv}{\phi}
\newcommand{\MSv}{\vartheta}
\newcommand{\MSm}{\pi_\vartheta}

\newcommand{\aMS}[1]{\qf{a}_{#1}}
\newcommand{\adMS}[1]{\qf{a}^\ast_{#1}}

\newcommand{\bT}[1]{\qf{b}_{#1}}
\newcommand{\bdT}[1]{\qf{b}^\ast_{#1}}

\newcommand{\ef}[1]{\xi_{#1}}

\newcommand{\mMS}{M_{\te{MS}}}
\newcommand{\mT}{M_{\te{T}}}
\newcommand{\mPhi}{m}

\newcommand{\Ls}{\te{s}}
\newcommand{\Lf}{\te{f}}

\newcommand{\Uf}[1]{\qf{1}_{#1}}
\newcommand{\Us}[1]{\qs{1}_{#1}}

\newcommand{\Hf}{\mathcal{H}_{\text{f}}}
\newcommand{\Hs}{\mathcal{H}_{\Ls}}

\newcommand{\pp}{\varepsilon}

\title{\it Backreaction in Cosmology}

\author[1]{S. Schander\thanks{sschander@perimeterinstitute.ca}}
\author[2]{T. Thiemann\thanks{thomas.thiemann@gravity.fau.de}}

\affil[1]{\small Perimeter Institute for Theoretical Physics,
31 Caroline Street North, Waterloo, ON, Canada N2L 2Y5}

\affil[2]{Institute for Quantum Gravity, Friedrich--Alexander--Universität Erlangen -- Nürnberg, \newline 
Staudtstr. 7, 91058 Erlangen, Germany}

\date{\small{\today}}

\begin{document}

\maketitle

\begin{abstract}
In this review, we investigate the question of backreaction in different approaches to cosmological perturbation theory, and with a special focus on quantum theoretical aspects. By backreaction, we refer here to the effects of matter field or cosmological inhomogeneities on the homogeneous dynamical background degrees of freedom of cosmology. We begin with an overview of classical cosmological backreaction which is ideally suited for physical situations in the late time Universe. We then proceed backwards in time, considering semiclassical approaches such as semiclassical or stochastic (semiclassical) gravity which take quantum effects of the perturbations into account. Finally, we review approaches to backreaction in quantum cosmology that should apply to the very early Universe where classical and semiclassical approximations break down. The main focus is on a recently proposed implementation of backreaction in quantum cosmology using a Born--Oppenheimer inspired method.  
\end{abstract}

\newpage

\renewcommand\contentsname{\it{Contents}}
\tableofcontents
\vspace{0.9cm}
\section{Introduction}
The $\Lambda$ cold dark matter ($\Lambda$CDM) concordance model, \parencite{CervantesCota2011,DodelsonSchmidt2021,DeruelleUzan2018}, based on the pillars of the Standard Model of particle physics and general relativity, has shaped our current view of the Universe, and has been the driving force behind many of the breakthroughs of modern cosmology, for example the prediction and the discovery of the cosmic microwave background radiation, \parencite{Gamov1948-01,Gamov1948-02,Alpher1948-01,Alpher1948-02,PenziasWilson1965,Planck2019I,Planck2020VI}. Modelled by only six parameters, \parencite{Planck2020VI,Spergel2015}, it features an impressive simplicity while correctly predicting and fitting most of the cosmological data, \parencite{Planck2019I,Planck2020VI}.
One of the most important assumptions within the $\Lambda$CDM paradigm is that the Universe is almost spatially homogeneous and isotropic, especially during its earliest phases, but even today when considered on its largest scales. The resulting simplification of Einstein's equations is remarkable as it reduces the ten coupled non--linear partial differential equations in four variables to two ordinary equations in one variable, with solutions known as the Friedmann--Lemaître--Robertson--Walker (FLRW) solutions, \parencite{Friedman1922,Friedman1924,Lemaitre1931,Robertson1933,Walker1937}.

Obviously, a look at the night sky reveals that the Universe is not homogeneous and isotropic, but is characterized by clusters of galaxies and stars, and large voids inbetween, \parencite{Zeldovich1982,Blumenthal1984,Colless2001,Cole2005,Ross2020}. For explanation, the concordance model assumes that smallest quantum fluctuations of the primordial matter and geometry have been stretched to the present time, thereby generating the observable large scale structure. Importantly, these inhomogeneities on any scale smaller than the observable Universe are presumed to evolve following the underlying FLRW background structure, but conversely their evolution does \emph{not} affect the global FLRW evolution. More precisely, it is assumed that effects from the small scale inhomogeneities onto the largest scales can be neglegted, i.e., there is no substantial backreaction. 

Doubts regarding the simplistic nature and the question of backreaction have gained momentum in recent years. In fact, the $\Lambda$CDM model, as appealing it may be, leads to the conclusion that approximately $69\%$ of the energy budget of our Universe consists of a yet unknown fluid, dubbed ``dark energy'', \parencite{Planck2020VI}, and which drives the very recent accelerated expansion of the Universe, \parencite{Riess1998,Perlmutter1999,Peebles2002}. Most of the remaining $31\%$ of the energy budget is credited to another yet unknown form of cold ``dark'' matter, \parencite{Peebles1982,Blumenthal1984,Planck2020VI}, which provides an explanation for the characteristic rotation and motion of the remaining $6\%$ of ordinary matter in the Universe. In summary, we are faced with the problem that we are literally in the dark about $94\%$ of the energy and matter content of the observable Universe. 

In recent years, these conceptual problems have been accompanied by important tensions in the estimates of certain cosmological parameters as made by different collaborations, \parencite{Pesce2020,DiValentino2020-02,DiValentino2020-03}. The evaluation of the Hubble constant $H_0$ as performed by the Planck collaboration (explicitely assuming a $\Lambda$CDM model) gives a value of $H_0 = (67.27 \pm 0.60) \text{km}/(\text{s}\cdot \text{Mpc})$, \parencite{Planck2020VI}, while the SH0ES collaboration finds $H_0 = (74.03 \pm 1.42) \text{km}/(\text{s}\cdot \text{Mpc})$, \parencite{Riess2019}, which in turn is based on the measurements of the Hubble Space Telescope. This leads to a tension at the $4.4 \sigma$ level, \parencite{DiValentino2020-02}. Furthermore, we point to the (albeit weaker) tensions regarding the measurement of the parameter $S_8$, a measure for the matter energy density $\Omega_m$ and the amplitude of structure growth $\sigma_8$, \parencite{Planck2020VI,DiValentino2020-03}.

On the other hand, the theoretical modeling of the early and very early Universe turns out to be a difficult undertaking, in particular the faithful consideration of all interactions within coupled quantum cosmological -- matter systems. Since classical cosmological perturbation theory and its various applications to the physics of our Universe, \parencite{Durrer2004,Mukhanov2005}, represents a successful formalism to model (most of) the cosmological data today, one of the most promising approaches to make progress in the field is to consider an inhomogeneous, but perturbative, \emph{quantum} cosmology, i.e., to establish a quantization of the well--known (possibly) gauge--invariant cosmological perturbation theories, \parencite{Brandenberger1993,Brandenberger2003,ElizagaNavascues2016}. In fact, there has been tremendous progress in developing such quantum cosmological perturbation theories, for example, in quantum geometrodynamics, \parencite{Brizuela2018,Kiefer2007}, in string cosmology, \parencite{Erdmenger2009}, as well as in loop quantum cosmology (LQC) and spinfoam cosmology, \parencite{Agullo2012,Cailleteau2011,ElizagaNavascues2016,Vidotto2010,Bianchi2010}, to mention but a few. Unfortunately, the majority of these approaches neglect backreaction effects from the inhomogeneous quantum fields on the homogeneous, dynamical degrees of freedom, or incorporate a series of assumptions which are hard to control, similar to the situation in classical cosmological perturbation theory.

It seems hence very timely to scrutinize and question the various assumptions of the concordance model of cosmology, and to develop suitable formalisms which are able to take interactions in coupled (quantum) cosmological models more realistically and unambigiously into account. In this review, we start by assessing the question of backreaction, i.e., whether cosmological inhomogeneities have an effect on the large scale evolution of the Universe, especially in view of the occurent inconsistencies within the standard model. We consider different aspects of backreaction, in particular, we discuss backreaction in classical, semiclassical and quantum mechanical models. Our main focus is on the purely quantum mechanical backreaction and we discuss one recent approach to including backreaction in quantum cosmology in more detail, \parencite{SchanderThiemannI}. The structure of the paper is then as follows.

In section \ref{sec:Classical Backreaction}, we provide an overview of the results in the field of \emph{classical} backreaction, which is particularly relevant for late time cosmological models. In section \ref{sec:Semiclassical Backreaction}, we consider \emph{semiclassical} backreaction which occurs when considering quantum fields on classical curved space times. Section \ref{sec:Quantum Backreaction} gives an overview of \emph{quantum} backreaction, i.e., backreaction that occurs in purely quantum theoretical models. In section \ref{sec:SAPT}, we focus on one particular approach to quantum backreaction which uses mathematical tools inspired by the Born--Oppenheimer approximation. Section \ref{sec:Discussion and Outlook} provides a final discussion and an outlook.

\section{Classical Backreaction}
\label{sec:Classical Backreaction}
Standard perturbative approaches to cosmological perturbation theory implicitely conjecture that backreaction, i.e., the effects of cosmological inhomogeneities on the global or macroscopic evolution of the Universe can be ignored. For purely classical models of the Universe that are particularly relevant for its late time evolution, this conjecture has generated an intense debate over the last decades. And still, there is no consensus on the question of backreaction in the classical regime, see for example the reviews by \textcite{Clarkson2011,Bolejko2016,Ellis2011}. 

The question of backreaction is closely related to the fitting problem, \parencite{Ellis1987}, and the problem of averaging, \parencite{Clarkson2011}. In fact, an intuitive way to access the effects of small inhomogeneities on the macroscopic scales is to construe an averaging procedure that defines new homogeneous variables by integrating the inhomogeneous fields over a certain space time domain, and to compare their properties and dynamics to the assumed FLRW Universe, \parencite{Ellis2011}. However, it is inadmissible to conclude from the validity of Einstein's equations for the inhomogeneous fields on the smallest scales (where they have been excellently checked), that the \emph{averaged} fields satisfy the Einstein equations, \parencite{Paranjape2012}. This is because evaluating the Einstein tensor and taking a space (time) average do not commute in general. Hence, the averaging procedure can lead to additional contributions to Einstein's equations that might be considered as effective source terms for the geometry, see for example, \parencite{Buchert2000,Buchert2001,Ellis2011,Paranjape2012}.

As it turns out the results regarding the form and strength of backreaction depend heavily on the averaging procedure and the matter model being chosen, as well as on the choice of space time volumes to be integrated over. In the non--perturbative regime, the two most discussed averaging procedures are the scalar averaging scheme by \textcite{Buchert2007} and the Macroscopic Gravity approach by \textcite{Zalaletdinov1997,Zalaletdinov2008}, and we refer the interested reader to the above cited reviews for extensive lists of further approaches. Buchert's scheme focuses on building spatial averages of scalar fields and derives effective (scalar) equations of motion for the averaged quantities, for example an improved Raychaudhuri equation for the averaged scale factor that includes a kinematical backreaction term. While being technically easier to implement, the Buchert scheme relies on a system of scalar equations that is not closed, \parencite{Clarkson2011,Ellis2011}, and consequently requires additional information to fix the solution. Besides, the averaging demands to fix suitable spatial domains and hence, a hypersurface slicing. In contrast, the Macroscopic Gravity approach is manifestly covariant but requires to define an auxiliary so--called bi--local transport operator, \parencite{Ellis2011}. Physical applications of these schemes yield a range of different results, ranging from explaining the recent accelerated expansion of the Universe or the $H_0$--tension, \parencite{Buchert2012,Heinesen2020}, to negligible backreaction effects, \parencite{Paranjape2007,Paranjape2008}.

The second issue concerns the question of how to appropriately model the matter content of the Universe. Many of the afore--mentioned approaches (among many others) assume the matter content to be modeled by a fluid, which is likely to be a poor approximation to the true lumpy late time Universe. Models with more realistic matter distributions are for example the Timescape Cosmology by \textcite{Wiltshire2009}, who separates the Universe into underdense expanding regions bounded by overdense virialized structures, the Swiss Cheese Model \parencite{Kantowski1969,Biswas2007,Tomita1999}, or modifications of FLRW--Universes that cut spherically symmetric Lemaître--Tolman--Bondi or Szekeres dust space time regions, \parencite{Bolejko2010,Marra2007}, to mention but a few. By construction, many of these models follow the evolution of an appropriately fitted FLRW model since they assume a background structure from the beginning. Consequently, they do not attack the backreaction problem outlined before. In contrast, the model by \textcite{Lindquist1957} assembles static Schwarzschild regions without relying on any background, and which has been further investigated by \textcite{Clifton2010,Clifton2009}. In both cases, the models provide insights into backreaction effects on light propagation, \parencite{Krasinski2010,Sussman2011}, which points to another important topic.  

In fact, cosmological observations such as the distance--redshift relation or the angular diameter distance rely on measurements of light, travelling along our past lightcone in a very inhomogeneous Universe. The seminal work by \textcite{Kristian1966} laid out the basis for analyzing backreaction on light propagation. \textcite{Flanagan2005}, for example, used these ideas to compute the deceleration parameter as measured by comoving observers. \textcite{Gasperini2011} define a covariant light--cone average for the backreaction problem, see also \parencite{Fanizza2019} for a more recent, generalized proposal. \textcite{Rasanen2009,Rasanen2010} derives a relationship of the redshift and the angular diameter distance to the average expansion rate for statistically homogeneous and isotropic universes, based on Buchert's approach, and \parencite{Barausse2005,Bonvin2005} evaluate the distance--redshift relation and the luminosity distance in a perturbative framework. Most recently, \textcite{Koksbang2019,Koksbang2020,Koksbang2021,Heinesen2021,HeinesenI2020} investigated the effects of inhomogeneities and averaging on a possible redshift drift.

Besides the non--perturbative approaches to backreaction discussed above, many approaches that attempt to make direct contact with cosmological observations restrict their analysis to cosmological perturbation theory in an FLRW Universe. Most of them consider flat $\Lambda$CDM models with Gaussian scalar perturbations as initital conditions, \parencite{Ellis2011}. To evaluate backreaction, they compute the deviations to the Hubble expansion rate or similar variables that are caused by backreaction, \parencite{Kolb2004,Li2007,Clarkson2009,Kolb2009,Brandenberger2018}, or give effective Friedmann equations with additional contributions, \parencite{Behrend2007,Baumann2010,Paranjape2007,Brown2008,Peebles2009}. The idea is to perform appropriate spatial averages of the perturbed quantities and to use the given statistical information of the perturbation fields in guise of their power spectra.

It turns out that due to the smallness of the gravitational potential and the power suppression of modes on large scales, backreaction for the expansion rate is always small. However, the backreaction to the deceleration parameter $q$ and the variance of the Hubble rate depend on an auxiliary UV--cutoff that might lead to large backreaction even if it is set by scales larger than the non--linearity scale, \parencite{Clarkson2011}. Other approaches to backreaction in the linear regime are \parencite{Baumann2010} who propose a reformulation of perturbation theory that leads to small backreaction on the largest scales but affects the baryon accoustic oscillations, and \parencite{GreenWald2010,GreenWald2011,GreenWald2013,GreenWald2014} who claim, using a point limit process, that backreaction can never mimic dark energy and put strong constraints on its strength. All these perturbative approaches rely of course on the assumption that the Universe, consisting of large voids between matter dominated regions, can be well approximated with Newtonian methods, \parencite{Ellis2011}.

We also point to the quite recent advent of numerical tools that allow to simulate increasingly realistic models of the Universe, including relativistic effects \parencite{Loffler2012,Mertens2015} and N--body simulations, \parencite{Adamek2016,Barrera2019}. Using the N--body relativistic code ``gevolution'', \textcite{Adamek2017} find that backreaction on the expansion rate in a $\Lambda$CDM and an Einstein--de Sitter Universe remains small if one chooses averaging volumes related to the Poisson gauge, while when choosing comoving gauge backreaction is of the order of $15\%$. Other works in this respect were done by \textcite{Macpherson2018}, who also claim that backreaction effects are small, however based on a fluid approximation which breaks down as soon as it comes to shell crossing.

Finally, let us also point to the consideration of backreaction from long wavelength modes in models of the \emph{early} Universe. In this respect, early contributions were notably made by \textcite{TsamisWoodard1993,Tsamis1996}, as well as by \textcite{AbramoBrandenbergerMukhanov1997,MukhanovAbramoBra1996}. The latter works pursue a gauge--invariant formulation of the backreaction problem associated with an effective long wavelength energy momentum tensor, and within a slow--roll inflationary scenario. \textcite{Unruh1998} subsequently examined the question of whether this effect is indeed locally measurable, and it was found that such backreaction effects (in single field inflationary theories) can be absorbed by a gauge transformation, \parencite{Abramo2001,Geshnizjani2002}. However, backreaction of such fluctuations becomes locally measurable after introducing an additional (subdominant) clock field, \parencite{Geshnizjani2003}. This approach was first extended by \textcite{Marozzi2012} based on the gauge--invariant formalism by \textcite{Finelli2011}, and secondly by \textcite{Brandenberger2018} beyond perturbation theory. Further contributions were made by \textcite{Losic2005,Losic2008} who support the idea that backreaction represents a real and measurable effect.
 
\section{Semiclassical Backreaction}
\label{sec:Semiclassical Backreaction}
For considerations of backreaction in models of the very early Universe, the standard model of cosmology suggests that (at least) the matter fields should be studied in a quantum mechanical framework. The implementation of such ideas can be realized via different paths, and we consider here the approaches of semiclassical gravity, \parencite{Ford2005,Wald1977,Wald1978} and stochastic (semiclassical) gravity, \parencite{Calzetta1986,Jordan1986,Jordan1987,Hu2008}. Both approaches rely on the framework of quantum field theory on curved space times (QFT on CST), \parencite{Fulling1989,BirrellDavies1982}, which itself takes the effects of the classical curved space times on the quantum matter fields into account but in general \emph{not} the backreaction effects of the quantum fields on the classical background.

In such approaches, the backreaction problem was first brought in by \textcite{Wald1977} who considered the backreaction from particle creaction on a gravitational field. The idea of the semiclassical program is to consistently define an improved set of Einstein field equations in which the expectation value of the quantum stress--energy tensor $\qf{T}_{\mu \nu}$ of the matter fields with respect to an appropriate quantum state of the matter fields $\omega$ appears as a source term,
\begin{equation}
R_{\mu \nu} + \frac{1}{2} g_{\mu \nu} R = 8 \pi G \omega(:\qf{T}_{\mu \nu}:),
\end{equation}
where the quantization with respect to the matter fields is expressed using bold letters and the dots indicate the normal ordering of the stress--energy tensor. The state $\omega$ should be considered here as a positive linear functional in the sense of the algebraic approach to quantum field theory (QFT), \parencite{Araki1999,Haag1992}. 

The first goal of semiclassical gravity is to define a procedure that leads to a meaningful expression for the expectation value of the stress--energy tensor. In fact, the latter depends on products of operator--valued distributions, even for the simple case of a real--valued Klein--Gordon field, and its expectation value is in general a divergent expression. \textcite{Wald1977} gave a set of axioms that are required to hold for a suitable renormalization scheme. Possible proposals are the Hadamard point--splitting method, \parencite{Brunetti1999,Hollands2001}, and the adiabatic regularization procedure, \parencite{Parker1974,FullingParker1974,FullingParkerHu1974}. In either scheme, the result of the regularization procedure is a set of modified ``semiclassical'' Einstein equations. These equations are substantially harder to solve than the original Einstein equations and many studies restrict to cases of conformally coupled matter to avoid problems regarding the well--posedness and the stability of the solutions, \parencite{Ford2005}. Caution is also required regarding the question of self--consistency of the backreaction effects, as has been discussed by \textcite{Flanagan1996}.

Many applications of the semiclassical gravity approach to early Universe cosmology have been considered. For example, \textcite{Fischetti1979} analyzed the backreaction effects from a conformally invariant matter field in an FLRW Universe with classical radiation, and found that the trace anomaly can soften the cosmological singularity, but not avoid it. Other works in this direction were done by \textcite{Anderson1983I,Anderson1984II,Anderson1985III}, who also considered the trace effects on the particle horizon. A well--known example of trace anomaly effects from semiclassical gravity is the \textcite{Starobinsky1980} cosmological model.

Another application of semiclassical gravity is the study of backreaction of particle creation on the dynamics of the early Universe, as already conceived by \textcite{Wald1977}. \textcite{Grishchuk1977} as well as \textcite{HuParker1977} considered the effect of gravitons around the Planck time in an FLRW Universe with a classical, isotropic fluid. The model leads to a timely non--local (i.e., history--dependent) backreaction effect, \parencite{HuVerdaguer2020}. Similar studies were performed for anisotropic FLRW Universes and it was shown that particle production due to the shear anisotropy isotropizes space time, \parencite{Zeldovich1971,HuParker1978}. Regarding the effects of particle creation in a spatially inhomogeneous but isotropic Universe, we refer to the work by \textcite{Campos1993}. 

We also point to more recent works by \textcite{Finelli2002,Finelli2004} who specifically consider a slow--roll (almost de Sitter) phase of the very early Universe and compute a(n adiabatically) renormalized energy momentum tensor of quantum inflaton, respectively cosmological scalar fluctuations. In case of the cosmological scalar perturbations, they find that the energy momentum tensor is characterized by a negative energy density which grows during inflation, and also that backreaction is not a mere gauge artifact.

Further contributions to the topic of semiclassical gravity for cosmological situations were notably made by \textcite{Eltzner2010,Pinamonti2010,Dappiaggi2010,Hack2013,Dappiaggi2008,Gottschalk2018}, to mention but a few. Most recently, \textcite{Pinamonti2013,Meda2020} have made progress on the definition of the semiclassical theory for general couplings by proving existence and uniqueness of solutions in flat cosmological space times with a massive quantum scalar field. The idea of relating backreaction effects to the decay of a cosmological constant has for example been promoted by \textcite{Dymnikova2001}. We also point to the work by \textcite{Matsui2019} (and references therein) who claim that the approach of semiclassical gravity is not appropriate to describe the early Universe.
 
The second approach to evaluating backreaction in semiclassical cosmology that we present here, denoted stochastic gravity, \parencite{HuVerdaguer2020}, creates a link to open system concepts and statistical features such as dissipation, fluctuations, noise and decoherence. It employs a so--called closed time path coarse grained effective action (CTP CGEA), \parencite{Jordan1986,Calzetta1986,Jordan1987}, in order to derive a set of modified semiclassical Einstein equations, denoted as Einstein--Langevin equations. It includes the semiclassical approach but extends it by a stochastic noise term, \parencite{Hu1994}.

Some of the first applications of the CTP CGEA formalism to the backreaction problem in cosmology were made by \textcite{Calzetta1986,Calzetta1989,Calzetta1993}. \textcite{Hu1994} derived the Einstein--Langevin equations for the case of a free massive scalar field in a flat FLRW background, as well as for a Bianchi Type--I Universe. The case of a massless conformally coupled field was discussed in \parencite{Campos1993}. The scope of works includes topics such as stochastic inflation, where quantum fluctuations present in the noise term backreact on the inflaton field, \parencite{CalzettaHu1995,Lombardo1996}, as well as studies of the reheating phase in inflationary cosmology, \parencite{Boyanovsky1995,Ramsey1997}. The formalism was also used by \textcite{SinhaHu1991} to check the validity of the minisuperspace approximation in quantum cosmology. 

We also point to one of the most prominent applications of stochastic methods to early Universe cosmology by \textcite{Starobinsky1982,Starobinsky1988}. His stochastic inflationary model evaluates backreaction of small scalar field quantum perturbations on the corresponding long wavelength modes (which are assumed to behave classically) by additional stochastic terms in the long wavelength equations of motion. A slow roll behavior of the background is assumed. Interestingly, it has been shown that the stochastic and the quantum field theoretic approaches to perturbations in the early Universe yield the same results, \parencite{Starobinsky1994,Tsamis2005,Finelli2008}. For recent considerations of stochastic inflation beyond the (strict) slow roll conditions, we refer to the work by \textcite{Pattison2019} and references therein.

Both approaches, semiclassical as well as stochastic gravity regard the gravitational field as a classical entity from the start while the matter fields are considered to be of quantum nature. While this represents a seminal progress to incorporating quantum effects of the matter fields in the early Universe, it can and should be questioned whether this somehow incompatible approach (classical and quantum fields treated at the same level) survives the test of future observations, and whether it should be replaced by a more consistent approach -- quantum gravity -- at least for the earliest moments of the cosmic history.

\section{Approaches to Quantum Backreaction}
\label{sec:Quantum Backreaction}
The question of backreaction in quantum gravity and quantum cosmology encompasses a variety of different approaches and definitions of backreaction. In quantum cosmology, backreaction is usually identified as the effects from the inhomogeneous quantum perturbation fields on the (quantum) homogeneous and isotropic degrees of freedom, which is also the notion of backreaction used in the next section, \parencite{SchanderThiemannIV}. This approach is tightly related to a perturbative expansion with respect to the inverse Planck mass $m_{\te{Pl}}^{-1} = (G/\hbar)^{1/2}$, where $G$ is Netwon's constant and we set $\hbar\equiv1$. More precisely, it employs a Born--Oppenheimer type scheme, \parencite{BornOppenheimer1927}, with respect to $m_{\te{Pl}}^{-1}$. We will thus focus on implementations of the Born--Oppenheimer method to quantum gravity and quantum cosmology.

In fact, the idea that quantum gravity can be considered as a perturbative theory with respect to $m_{\te{Pl}}$ was already introduced by \textcite{Brout1987}. The first investigations of backreaction in quantum gravity that rely on this expansion were performed in the framework of quantum geometrodynamics, \parencite{Wheeler1957}, see also \parencite{Kiefer2007} for an extensive overview. The idea is to expand the Wheeler--DeWitt equation in terms of the ratio of the Planck mass and the matter field mass, \parencite{Kiefer1991}. A different idea, conceptually similar to the schemes considered here, is to use a Born--Oppenheimer type approach, relying on the same perturbation parameter. Different considerations of the problem (giving rise to similar results) can be found in the works by \textcite{Brout1989,Kiefer1994,Bertoni1996}, (again, for a summary, see \parencite{Kiefer2007}). A review of the ideas of Born and Oppenheimer will be given in the next section, but to understand its use in the given context we present the key ideas. 

For simplicity, let $Q$ denote the gravitational and $q$ the matter degrees of freedom. The Born--Oppenheimer scheme employs an ansatz solution for the quantum Hamiltonian and momentum constraint of the form, \parencite{Kiefer2007},
\begin{equation}
\Psi(q,Q) = \sum_n \chi_n(Q) \psi_n(q,Q),
\end{equation}
where $\lbrace \psi_n(q,Q) \rbrace_n$ is supposed to be a known orthonormal basis of the matter Hilbert space that solves the matter part of the constraint and $Q$ is to be considered as an external parameter for this eigenvalue problem. Then, one applies the constraints to $\Psi$ and applies some $\psi_k(q,Q)$ from the left (i.e., one considers the inner product of the matter states). This gives rise to constraint equations for the geometric factors $\chi_n(Q)$, which can be seen as an effective quantum problem for the geometric part, including the backreaction effects of the quantum matter system. In this scenario, the Born--Oppenheimer approximation consists in neglecting the contributions that enter with higher orders in $m_{\te{Pl}}^{-1}$.

In order to extract physical results from the formalism, one can additionally employ a semiclassical approximation (which is however \emph{independent} of the Born--Oppenheimer approach). This should yield the semiclassical limit of quantum gravity, i.e., a matter QFT on CST. It is common to employ a WKB ansatz for the geometrical states $\chi_n(Q)$ of the form,
\begin{equation}
\chi_n(Q) = C_n(Q) e^{i m_{\text{Pl}}^2 S[Q]},
\end{equation}
where $S[Q]$ stands for the geometric action in the geometrodynamical approach. The perturbative scheme in $m_{\text{Pl}}^{-1}$ eventually yields the semiclassical Einstein equations. In this sense, these approaches evaluate the backreaction of the quantum matter fields on the quantum or classical geometry. 

One can apply the Born--Oppenheimer and WKB approximations in a different manner. Instead of taking the expectation value with respect to the quantum matter system, one applies the Wheeler--DeWitt constraints on the total Born--Oppenheimer ansatz function and uses the WKB approximation for the geometrical part. Restricting again to the lowest order with respect to the inverse Planck mass, this yields a quantum constraint for the matter wave function which depends on the classical action (through the WKB ansatz), and derivatives with respect to the spatial metric thereof. The idea of the above--cited works (and also of \parencite{Briggs2000}) is to introduce an external time parameter that depends on this derivative, hence giving rise to a Schrödinger equation for the matter system that includes the backreaction of the geometry through the geometry--dependent time derivative. In fact, this gives rise to a notion of time in a formerly background independent framework. Such ideas go back to \textcite{DeWitt1967} and have been applied to a variety of cosmological situations (see \parencite{Kiefer2007} and references therein). It is however a different notion of backreaction than the one considered in the next section. Besides, the present approach uses a Born--Oppenheimer approach \emph{plus} a semiclassical WKB approximation while the next section uses the purely quantum mechanical space adiabatic perturbation extension of the Born--Oppenheimer scheme. Applications of the former works to the inflationary paradigm with perturbations and a discussion of the question of unitary evolution of the perturbations can be found in the work by \textcite{Chataignier2020} and references therein. They also consider cosmological perturbations that include gravitational contributions (i.e., the Mukhanov--Sasaki variables). Similar approaches that do not split the system into geometric and matter parts but include (perturbative) parts of the gravitational degrees of freedom in the fast subsystem and (homogeneous) matter parts in the slow sector were already presented by \textcite{HalliwellHawking1987,Vilenkin1989}. This choice is also used in \parencite{SchanderThiemannIV}. 

The Born--Oppenheimer approximation was also considered within approaches to quantum gravity with other variable choices. \textcite{Giesel2009} aimed at an application of the Born--Oppenheimer methods to loop quantum gravity (LQG), \parencite{Thiemann2008,Rovelli2010}, using holonomy--flux variables or connection--flux variables. As it turns out, this choice of variables prevents the use of the Born--Oppenheimer methods since the flux operators are mutually non--commuting (which is a prerequisite for the Born--Oppenheimer scheme). Instead, they use commuting co--triad variables for the gravity sector and a scalar field for the matter sector to derive the semiclassical Einstein equations that take the backreaction of the quantum matter fields via an expectation value into account. \textcite{Giesel2009} consider their model on a discrete lattice (as it is common practice for approaches to LQG), and thus formally obtain a lattice QFT on a discrete curved space time. They also point to the possibility of pursuing the formal Born--Oppenheimer scheme and computing \emph{quantum} solutions to the gravity sector with the effective backreaction of the quantum matter fields. Besides, they introduce a hybrid approach (similar to the models we consider here) where the gravitational sector is restricted to FLRW solutions and the fast part of the system is given by the matter quantum fields. They also propose to introduce coherent states for the gravitational subsystem in order to make progress in finding solutions. Due to the complexity of the gravity--matter systems, the focus of this work lies on spelling out the conceptual ideas rather than technically carrying out the programme in detail.

More recently, \textcite{StottmeisterThiemann2016a,StottmeisterThiemann2016b,StottmeisterThiemann2016c} considered similar questions in the context of LQG but employed the more general scheme of space adiabatic perturbation theory (SAPT), \parencite{PST03}. Since in the latter approach, the variables of the slow, gravitational sector are not required to commute, it is in principle possible to apply the Born--Oppenheimer ideas also to LQG and related theories. The concrete implementation turns however out to be difficult due to the particular structure of the LQG phase space (which relies on a cotangent bundle of a compact Lie group rather than on a vector space) and its quantum representation. Other open issues of their attempts are related to the underlying graph structure of LQG models and the projective limits of finite dimensional truncations of the gravitational phase space that are needed in order to construct a continuum theory, \parencite{StottmeisterThiemann2016c}. They also point out that a major obstruction to the derivation of a QFT on CST from LQG lies in the inequivalent representations of quantum fields for different gravitational configurations, \parencite{StottmeisterThiemann2016c}. This problem is a generic feature of background dependent quantum field theories. In this work, we present a (perturbative) solution to this problem which makes the application of space adiabatic methods to quantum cosmology possible, \parencite{Fer+12,CMM15,CMM16}. 

For completeness, we also mention the application of Born--Oppenheimer methods within the spinfoam approach to LQC, \parencite{Rovelli2008}, (see \parencite{AsthekarBojowaldLewandowski2003,Bojowald2008} for the LQC approach), and that \textcite{CMM16} consider a conceptually different kind of Born--Oppenheimer approximation in the hybrid approach to LQC. These seminal works make important progress by first considering the problem of backreaction, but must either remain on a rather formal level or include various assumptions which are hard to control. We therefore advocate to employ the SAPT scheme presented in the next section which serves as an unambigious, rigorous and perturbative approach, in principle applicable to any quantum cosmological framework and realizable up to any perturbative order, to taking quantum cosmological backreactions thoroughly into account. The approach applies to a much wider variety of quantum systems in comparison to the Born--Oppenheimer approach (in particular to quantum cosmological perturbation theory), does not rely on the introduction of semiclassical ansatz states and iteratively provides quantum constraints or equations of motion whose solutions approximate the true solutions up to, in principle, indefinitely small errors.

\section{Quantum Backreaction with Space Adiabatic Methods}
\label{sec:SAPT}
Computing and including backreaction in (perturbative) quantum cosmology requires an approximation scheme that ideally takes the physical characteristics of the system into account. SAPT as proposed by \textcite{PST03} and extensions thereof are ideally suited to achieve this goal and to integrate backreaction effects into quantum cosmology, \parencite{SchanderThiemannI,NeuserSchanderThiemannII,SchanderThiemannIII,SchanderThiemannIV}. 

SAPT is a generalization of the well--known Born--Oppenheimer approximation for non--relati\-vistic molecular systems. Both approaches exploit the small ratio of two internal parameters such as the mass ratio of electrons and nuclei in a molecule to define a perturbation parameter,
\begin{equation}
\pp^2 := \frac{m_{\text{e}}}{m_{\text{n}}} \approx 5.46 \times 10^{-4} \ll 1,
\end{equation}
with the electron mass $m_{\text{e}} \approx 9.11 \times 10^{-31}$ kg and the nuclei mass $m_{\text{n}} \approx 1.67 \times 10^{-27}$ kg. In the simplest atom with one nucleus and one electron (although exact solutions are known for this case), the Hamilton function has the form,
\begin{equation}
H(q,P;x,y) = \frac{\pp^2 P^2}{2 m_{\text{e}}} + \frac{y^2}{2 m_{\text{e}}} + V(q;x),
\end{equation}
where $(q,P)$ and $(x,y)$ are the canonically conjugate pairs of the nucleus and the electron respectively. $V(q;x)$ is a smooth potential, typically a Coulomb potential depending on the distance between nucleus and electron. In this molecular set up, the equipartition theorem states that the kinetic energies of nuclei and electrons are of the same order, and hence, on average, the nuclei move much slower than the electrons with correspondent statistically--averaged velocities, $\langle v_{\text{n}} \rangle \approx \pp \langle v_{\text{e}} \rangle$. Born and Oppenheimer used this fact to define suitable ansatz solutions for the quantum mechanical problem: On the typical electronic time scale, the nuclei are at rest and the non--trivial electronic contributions of the Hamilton operator can be considered at fixed nuclei positions $q \in \mathbb{R}$,
\begin{equation}
\qf{H}_{\text{e}}(q) := \frac{\qf{y}^2}{2 m_{\text{e}}} + \qf{V}(q,\qf{x}),
\end{equation}
where the bold letters $\qf{y}$, $\qf{x}$ denote momentum and position operators of the electron defined on their respectively dense domains in $L^2(\mathbb{R},\mathrm{d}x)$. Ideally, the operator function $\qf{H}_{\te{e}}(q)$ admits a solvable $q$--dependent eigenvalue problem,
\begin{equation}
\qf{H}_{\te{e}}(q)\,\xi_n(q) = E_n(q)\, \xi_n(q),
\end{equation}
with a discrete $q$--dependent eigenbasis $\lbrace \xi_n(q) \rbrace_{n \in \mathbb{N}}$ in the \emph{fast} Hilbert space $\Hf := L^2(\mathbb{R},\mathrm{d}x)$, for which the so--called electronic energy bands $E_n(q)$ are gapped functions, i.e., $E_n(q) - E_m(q) \neq 0$ pointwise for $m\neq n$. One can use this eigenbasis as an ansatz solution for the full Hamilton operator $\qs{\qf{H}}$ (the Weyl quantization with respect to the nuclei sector is labeled by hats), and ask whether it provides an approximate solution to the entire problem. Equivalently, we can define for every electronic eigensolution the direct integral operator,
\begin{equation} \label{eq:Direct Integral Operator}
\qs{\qf{\pi}}_n := \int_{\mathbb{R}}^\oplus \mathrm{d}q \,\xi_n(q) \langle \xi_n(q),\cdot \rangle_{\te{e}} = \int_{\mathbb{R}}^\oplus \mathrm{d}q \,\qf{\pi}_n(q),
\end{equation}
on the total Hilbert space $\mathcal{H} = L^2(\mathbb{R},\mathrm{d}q) \otimes \Hf$ and ask whether it commutes with $\qs{\qf{H}}$. Of course, the answer is in the negative, but it turns out that the commutator scales like $\pp$,
\begin{equation} \label{eq:Commutator BOA}
\left[ \qs{\qf{H}}, \qs{\qf{\pi}}_n \right] \sim \pp.
\end{equation}
This is because of the adiabatic relation between the electrons and the nuclei. By construction, the Weyl quantization $\qs{\qf{H}}_{\text{e}}(\qs{q})$ of the electronic Hamiltonian and $\qs{\qf{\pi}}_n$ commute. However, the remaining contribution to the Hamilton operator $\qs{\qf{H}}$, in particular the kinetic energy of the nucleus, scales like $\pp^2$ and leads hence to the estimate in equation \eqref{eq:Commutator BOA}. The Born--Oppenheimer approximation builds on this result and proposes to use the ansatz functions,
\begin{equation} \label{eq:BOA Ansatz}
\Psi(q;x) = \sum_n \psi_n(q) \xi_n(q;x),
\end{equation}
to solve the full quantum problem. In its simplest version, the scheme neglects any of the contributions that arise from applying the kinetic energy operator of the nucleus to the electronic ansatz functions (as they enter with small $\pp$--factors), and thus results in an effective eigenvalue problem for the nucleus only,
\begin{equation}
\left( \frac{\qs{P}^2}{2 m_{\text{n}}} + E_n(\qs{q}) \right) \psi_n(q) \equiv E \psi_n(q).
\end{equation}
If this nucleonic eigenproblem can be solved, the scheme leads in fact to viable results for the stationary energy spectra of molecules, which are given by the energy solutions $E$, and which include the backreaction of the electrons via the potential energy $E_n(\hat{q})$. 

Unfortunately, the Born--Oppenheimer approach comes with some limitations which preclude its application to more complicated systems. Firstly, the scheme explicitely uses that the electronic eigenfunctions depend only on the \emph{configuration} variable of the nucleus. Was the coupling between electrons and nuclei provided by non--commuting slow operators, for example by $\qs{q}$ \emph{and} $\qs{p}$, the scheme would fail since the direct integral construction in equation \eqref{eq:Direct Integral Operator} builds on the commutativity (i.e., the existence of a common spectrum) of the coupling operators. Also one could not define the ansatz functions in equation \eqref{eq:BOA Ansatz}. Secondly, the scheme does not provide a simple extension to better error estimates. This becomes problematic if one is interested in the dynamical evolution of the system. The interesting dynamics of the nuclei happens on time scales $t_{\te{n}} \sim t_{\te{e}}/\pp$ or larger, but considering the evolution generated by $\qs{\qf{H}}$ with respect to the above ansatz functions, the scheme cannot lead to trustworthy results due to the commutator relation \eqref{eq:Commutator BOA}.

It should be possible to do better. In fact, the adiabatic theorem \parencite{Teu03} states that under certain conditions (to be discussed in the sequel), there exists an orthogonal projection operator $\qs{\qf{\Pi}} \in \mathcal{B}(\mathcal{H})$ in the bounded operators on the total Hilbert space $\mathcal{H}$ such that,
\begin{equation}
\left[ \qs{\qf{H}}, \qs{\qf{\Pi}} \right] = \mathcal{O}_0( \pp^\infty),
\end{equation}
where the right hand side means that for all $m \in \mathbb{N}$, there exists a constant $C_m < \infty$ such that $||[ \qs{\qf{H}}, \qs{\qf{\Pi}}]||_{\mathcal{B}(\mathcal{H})} \leq C_m \pp^m$, in the norm of bounded operators on $\mathcal{H}$. Most importantly, $\qs{\qf{\Pi}}$ can be constructed by a ``semiclassical symbol'' function, i.e., an operator--valued ansatz function like the almost--projector function $\qf{\pi}_n \in C^\infty(\mathbb{R},\mathcal{B}(\Hf))$ from the simple example above. This symbol function appears as an asymptotic series in the perturbation parameter $\pp$, and -- to anticipate the result -- the equivalent of $\qf{\pi}_n(q)$ will serve as the base clause to an iterative $\pp$--scheme to compute better and better approximations to $\qs{\qf{\Pi}}$. This is the idea of SAPT, \parencite{PST03}.

SAPT uses an $\pp$--scaled phase space (or deformation) quantization scheme, \parencite{BlaszakDomanski2012} for the slow subsector of the system, while retaining a standard Hilbert space representation for the fast sector. Phase space quantum mechanics is a formulation of quantum mechanics that employs an algebra of phase space functions $\mathcal{A}_Q$ instead of using the standard operator algebra in the Hilbert space representation of quantum mechanics. Quantum mechanical observables are thus represented by real--valued phase space functions. The pullback of the operator product to the phase space algebra gives rise to a non--commutative star product $\star$. Since the star product reduces to the commutative multiplication of phase space functions in the limit $\hbar \rightarrow 0$, this formulation of quantum mechanics is also known as deformation quantization. The standard textbooks by \textcite{Fol89,Hoermander1985a,Hoermander1985b,DimassiSjoestrand1999} give thorough introductions to the usual \emph{scalar}--valued phase space quantization scheme and pseudodifferential calculus, but the situation here is more subtle. SAPT requires to consider \emph{operator}--valued symbol functions, in particular functions on the slow phase space with values in the operators on the fast Hilbert space. We will thus deal with symbol functions of the form $\qf{A}(q,p) \in C^\infty(\Gamma_{\text{s}}, \mathcal{B}(\mathcal{H}_{\text{f}}))$ where $\Gamma_{\text{s}}$ denotes the slow phase space, \parencite[Appendix A]{Teu03}. It is straightforward to map the symbol functions (operator--valued or not) to their operator representatives in the standard Hilbert space approach. The concrete prescription depends of course on the operator ordering that one chooses. In case of the symmetric Weyl quantization prescription, this relation is provided by the Weyl correspondence, \parencite{DHS80}, and a symbol function appears as the kernel of an integral operator that acts on an element of the Hilbert space, \parencite{Teu03}.

Symbol functions which give rise to admissible operators in the quantum theory can be classified by their asymptotic behavior on the slow phase space. One important class of symbols relevant for SAPT are the semiclassical symbols $S^m_\rho$, with $m\in \mathbb{R}$ and $0 \geq \rho \geq 1$. An operator--valued function $\qf{A} \in C^\infty(\mathbb{R}^2,\mathcal{B}(\Hf))$ is in the symbol class $S^m_\rho(\mathcal{B}(\Hf))$ if for every $\alpha$, $\beta \in \mathbb{N}$, there exists a positive constant $C_{\alpha,\beta}$ such that, 
\begin{equation}
\sup_{q \in \mathbb{R}} \left\lVert \left( \partial_q^\alpha \partial_p^\beta \qf{A} \right)(q,p) \right\rVert_{\mathcal{B}(\mathcal{H}_{\te{f}})} \leq C_{\alpha,\beta} \langle p \rangle^{m-\rho|\beta|},
\end{equation}
for every $p\in \mathbb{R}$, and $\langle p \rangle = (1+|p|^2)^{1/2}$, \parencite{Teu03}. For such symbols the Weyl ordering prescription for quantum theory gives rise to a specific star product, and we can finally make sense of the space adiabatic perturbation idea. For two such operator--valued symbols on the slow phase space $\qf{A} \in S^{m_1}_\rho(\mathcal{B}(\Hf))$, $\qf{B}\in S^{m_2}_\rho(\mathcal{B}(\Hf))$, their star product is given by,
\begin{equation}
(\qf{A} \star \qf{B})(q,P) = \left.\exp \left( \frac{i \hbar}{2} \left( \partial_x \partial_P - \partial_q \partial_D \right) \right) \qf{A}(x,D)\, \qf{B}(q,P) \right|_{x=q,D=P} ~ \in S^{m_1+m_2}_\rho(\mathcal{B}(\Hf)).
\end{equation}
We note that the exponential has a series expansion which could be considered as a series with respect to $\hbar$ in the given context. Alluding to the adiabaticity of the system, it is reasonable to define a rescaled momentum operator $p:= \pp P$. Replacing $P$ and $D$ by their $\pp$--scaled versions in the star product formula, the new expansion parameter is $\pp\hbar$. Any star product of symbol functions can thus be written in a series expansion in $\pp$. Comparing terms of the same polynomial order in $\pp$, this defines a perturbation theory for quantum mechanical equations which will iteratively solve the eigenvalue problems of interest. The first two orders of the rescaled star product are given by,
\begin{equation}
(\qf{A} \star \qf{B})(q,p) = (\qf{A}_0\cdot \qf{B}_0)(q,p) + \frac{i \pp \hbar}{2} \lbrace \qf{A}_0(q,p), \qf{B}_0(q,p) \rbrace + \pp (\qf{A}_0\cdot \qf{B}_1 + \qf{A}_1\cdot \qf{B}_0 )(q,p) + \mathcal{O}(\pp^2),
\end{equation}
where the symbol functions have been expanded with respect to $\pp$ according to $\qf{A} = \sum_k \pp^k \qf{A}_k$ and $\qf{B} = \sum_k \pp^k \qf{B}_k$. Note that the $\pp$--rescaling changes the whole symplectic structure (we now have, $\lbrace q, p \rbrace_{\text{s}} = \pp$), as well as the canonical commutation relations since we obtain,
\begin{equation}
\left[ \qs{q},\qs{p} \right]_{\text{s}} = i \pp \hbar\, \qs{1}_{\text{s}}.
\end{equation}
From now on, we will set $\hbar \equiv 1$. Note that in the original Born--Oppenheimer approximation, the perturbative parameter occurs (after an appropriate rescaling) only in the Hamiltonian and the perturbation theory consists in splitting the Hamiltonian (and its spectrum) accordingly, while \emph{here} the quantum algebra is redefined, giving rise to a(n in principle infinite) perturbation series in $\pp$.  
\subsection{Space Adiabatic Perturbation Theory}
SAPT as introduced by \textcite{PST03} places a set of conditions on the physical system under consideration. These are, in some respects, quite restrictive. However, if one accepts to abandon certain results, such as the convergence of the perturbative series, it is possible to milden the conditions.  Here, we present the original conditions introduced by Panati, Spohn and Teufel for a system with $d$ slow and $k$ fast degrees of freedom, and which can be split into four categories:
\begin{enumerate}
\item[(C1)] \textit{The state space} of the system decomposes as,
\begin{equation}
\mathcal{H} = L^2(\mathbb{R}^d) \otimes \Hf = L^2(\mathbb{R}^d, \Hf),
\end{equation}
where $L^2(\mathbb{R}^d)$ is the state space of the system whose rate of change is by a factor $\pp^l$, $l \in \mathbb{R}^+$, smaller than the rate of change of the (environmental) system $\Hf$. The latter is assumed to be a separable Hilbert space.
\item[(C2)] \textit{The quantum Hamiltonian} $\qs{\qf{H}}$, (may it be an operator or a constraint) is given as the Weyl quantization of a semiclassical symbol $\qf{H} \in S^m_{\rho}(\pp, \mathcal{B}(\Hf))$, i.e., $\qf{H}$ asymptotically approaches an $\pp$--series,
\begin{equation}
\qf{H}(\pp,z) \asymp \sum_{j=0}^{\infty} \pp^j \qf{H}_j(z),
\end{equation}
where $\qf{H}_j \in S^{m-j \rho}_\rho (\mathcal{B}(\Hf))$ for all $j\in \mathbb{N}$ and $z:=(q,P) \in \mathbb{R}^{2d}$. The appropriate notion of convergence is provided by a Fréchet semi--norm in $S^m_\rho(\mathcal{B}(\Hf))$, see \parencite{Teu03} for further details.
\item[(C3)] For any fixed $z\in\mathbb{R}^{2d}$, the spectrum $\sigma(z)$ of the principal symbol $\qf{H}_{0}(z)$ of $\qf{H}(\pp,z)$ has isolated parts $\sigma_n(z)$, $n\in \mathbb{N}$. Picking one such $\nu\in \mathbb{N}$ and therefore suppressing any $n$--dependence in the following, the minimal distance between the elements of $\sigma_\nu(z)$ and the remainder of the spectrum $\sigma_{\te{rem}}(z) := \sigma(z) \backslash \sigma_\nu(z)$ displays a non--vanishing gap. According to its characteristics with varying $z$, the gap can be classified by means of a parameter $\gamma$. \\
\textit{Conditions} $(\textit{Gap})_{\gamma}$ : Let $f_\pm \in C^0(\mathbb{R}^{2d}, \mathbb{R})$ be two continuous functions with $f_- \leq f_+$. 
\begin{itemize}
\item[(G1)] \textit{Enclosing interval.} For every $z\in\mathbb{R}^{2d}$ the isolated part of the spectrum $\sigma_n(z)$ is entirely contained in the interval $I(z):=[f_-(z), f_+(z) ]$.
\item[(G2)] \textit{Gap to the remainder.} The distance between the remainder of the spectrum, $\sigma_{\te{rem}}(z)$ and the enclosing interval $I(z)$ is strictly bigger than zero and increasing for large momenta, i.e.,
\begin{equation}
\text{Dist} \left[ \sigma_{\te{rem}}(z), I(z)\right] \geq C_{\te{g}} (1+p^2)^{\frac{\gamma}{2}}. 
\end{equation}
\item[(G3)] \textit{Boundedness of the interval.} The width of the interval $I(z)$ is uniformly bounded, \textit{i.e.},
\begin{equation}
\sup\limits_{z\in\mathbb{R}^{2d}} \left| f_+(z)- f_-(z) \right| \leq C_{\te{d}} < \infty.
\end{equation}
\end{itemize}
\item[(C4)] \textit{Convergence Condition}. If the system satisfies the gap condition $(\te{C3})_\gamma$ for some $\gamma \in \mathbb{R}$, the Hamilton symbol $\qf{H}$ must be in $S^\gamma_\rho$. If $\rho = 0$, also $\gamma$ must vanish. If $\rho>0$, $\gamma$ can be any real number but $\qs{\qf{H}}$ must be essentially self--adjoint on $\mathcal{S}(\mathbb{R}^d,\Hf)$.
\end{enumerate}
Condition (C4) is not vital in order to perform the formal computations in the following. It ensures however that for considerations on the whole slow phase space $\Gamma_{\te{s}}$, the error estimates of SAPT are bounded everywhere on $\Gamma_{\te{s}}$. In particular, the adiabatic decoupling is said to be \emph{uniform}. Note also that the requirement that $\qf{H}$ has values in the \emph{bounded} operators is violated for many physical systems of interest. In such cases, the space adiabatic scheme cannot be immediately applied and the convergence of the perturbative expansion has to be examined by independent methods, \parencite{PST03}.

Given the conditions (C1) -- (C4), the space adiabatic theorem introduces a perturbative construction scheme that is based on iteratively computing three symbol functions: the Moyal projector $\qf{\pi} \in S^0_\rho$, the Moyal unitary $\qf{u} \in S^0_\rho$ and an effective Hamiltonian $\qf{H}_{\te{eff}} \in S^m_\rho$. The Moyal projector serves to identify a subspace of the total Hilbert space which is almost invariant under the dynamics of $\qs{\qf{H}}$ and which is associated with one particular quantum number $\nu \in \mathbb{N}$ of the fast sector. The Moyal unitary $\qf{u}$ is an auxiliary structure which gives rise to a unitary operator that maps the relevant subspace to a much simpler reference subspace. In fact, the original subspace is a technically complicated object and cannot provide us with a simple procedure to derive the (approximated) dynamics in the subspace. The reference subspace is trivial with respect to the fast subsystem and allows to compute the dynamics of the slow sector including the backreaction of the fast degree(s) of freedom. It is used to derive an effective Hamiltonian symbol $\qf{H}_{\te{eff}}$ whose solutions are approximate solutions to the full Hamilton operator $\qs{\qf{H}}$. More precisely, \parencite{Teu03},
\begin{enumerate}
  \item[(S1)] There exists a unique formal symbol, $\qf{\pi} = \sum_{i\geq0} \pp^i \qf{\pi}_i$, with $\qf{\pi}_i \in S^{-i\rho}_{\rho}(\mathcal{B}(\mathcal{H}_{\text{f}}))$, such that $\qf{\pi}_0$ is the spectral projection of $\qf{H}(q,p)$ corresponding to $\sigma_\nu (q,p)$ and with the properties,
  \begin{equation*}
    \text{(S1--1)}~~ \qf{\pi} \star_\pp \qf{\pi} = \qf{\pi},~~~~~~~~~   \text{(S1--2)}~~  \qf{\pi}^{\ast} = \qf{\pi},~~~~~~~~~ \text{(S1--3)}~~  \qf{H} \star_\pp \qf{\pi} - \qf{\pi} \star_\pp \qf{H} = 0.
  \end{equation*}
  It can be shown that the Weyl quantization of a resummation of $\qf{\pi}$, which we denote by $\qf{\pi}_{\pp}$ is $\mathcal{O}_0(\pp^{\infty})$--close to an operator $\qs{\qf{\Pi}}$, i.e., $\qs{\qf{\Pi}} = \qs{\qf{\pi}}_{\pp} + \mathcal{O}_0(\pp^{\infty})$ and that, $[\qs{\qf{H}},\qs{\qf{\Pi}}] = \mathcal{O}_0(\pp^{\infty})$. 
  \item[(S2)] Let $\qf{\pi}_{\te{R}}$ be the projection on some reference subspace $\mathcal{K}_{\text{f}} \subseteq \Hf$. Assuming that there exists a symbol $\qf{u}_{0} \in S^0_{\rho}(\mathcal{B}(\mathcal{H}_{\text{f}}))$, such that, $\qf{u}_{0}\cdot\qf{\pi}_{0}\cdot\qf{u}_{0}^{\ast} = \qf{\pi}_{\te{R}}$, then there is a formal symbol $\qf{u} = \sum_{i\geq0} \pp^i \qf{u}_i$ with $\qf{u}_i \in S^{-i\rho}_{\rho}(\mathcal{B}(\mathcal{H}_{\text{f}}))$ such that,
  \begin{equation*}
    \text{(S2--1)}~~ \qf{u}^{\ast} \star_\pp \qf{u} = \qf{1},~~~~~~~~~   \text{(S2--2)}~~  \qf{u} \star_\pp \qf{u}^{\ast} = \qf{1},~~~~~~~~~ \text{(S2--3)}~~  \qf{u} \star_\pp \qf{\pi} \star_\pp \qf{u}^{\ast} = \qf{\pi}_{\te{R}}.
  \end{equation*}
  The Weyl quantization of a resummation of $\qf{u}$, which denoted by $\qf{u}_\pp$ gives rise to an operator, $\qs{\qf{U}} = \qs{\qf{u}}_\pp + \mathcal{O}_0(\pp^{\infty})$ for which it holds true that, $\qs{\qf{U}}\,\qs{\qf{\Pi}}\,\qs{\qf{U}} = \qs{\qf{\pi}}_{\te{R}}$.
  \item[(S3)] There exists a formal, effective Hamilton symbol $\qf{h}_{\text{eff}} = \sum_{i\geq0} \pp^i \qf{h}_{\text{eff},i}$ defined as,
  \begin{equation*}
    \qf{h}_{\text{eff}} := \qf{u} \star_\pp \qf{H} \star_\pp \qf{u}^{\ast}.
  \end{equation*}
  For systems with an external time parameter $t$ and the Weyl quantizations $\qs{\qf{u}}$ and $\qs{\qf{h}}_{\text{eff}}$, we have, 
  \begin{equation}
  e^{-i\, \qs{\qf{H}}\,s} - \qs{\qf{u}}^\dagger e^{-i\, \qs{\qf{h}}_{\text{eff}}\, s}\, \qs{\qf{u}} = \mathcal{O}_0(\pp^{\infty}|s|).
  \end{equation}
\end{enumerate}
In the next section, we will make these formal definitions and results by \textcite{PST03} more explicit and apply the space adiabatic scheme to a simple cosmological model up to second order in the perturbations.

\subsection{Backreaction in Quantum Cosmology}
As an illustrative example for the space adiabatic scheme, let us consider Einstein general relativity, reduced to spatial homogeneity and isotropy, including a cosmological constant $\Lambda >0$, and coupled to a spatially homogeneous, isotropic and real Klein--Gordon field $\phi_0$ with mass $m>0$ and coupling constant $\lambda \in \mathbb{R}$. This section is based on \parencite{NeuserSchanderThiemannII} to which we refer for more details. We assume a globally hyperbolic space time manifold and a metric with Lorentzian signature $(-,+,+,+)$. The only dynamical degree of freedom of the metric is the scale factor $a \geq 0$. The lapse function $N$ is a Lagrange multiplier and will be fixed to $N \equiv 1$. We perform a $(3+1)$--split of the manifold into space and time which admits spatial  hypersurfaces $\sigma$ which we fix to be compact, flat three--tori $\mathbb{T}^3$ with side lengths $l\equiv 1$. The cosmological action is,
\begin{equation}
S\left[a(t),\phi_0(t)\right] = \int_{\mathbb{R}} \mathrm{d}t \left( - \frac{1}{\kappa} \left(3 \dot{a}^2 a + \Lambda a^3 \right) + \frac{1}{2\lambda} \left( \dot{\phi}_0^2 - m^2 \phi_0^2 \right) \right),
\end{equation}
where $\kappa = 8\pi G$ and $\lambda$ are the gravitational and scalar field coupling constants. If both, $(a,\phi_0)$ are dimensionless, as we assume, then both coupling constants have the same dimension, and we define the dimensionless ratio,
\begin{equation}
\pp^2 := \frac{\kappa}{\lambda}.
\end{equation}
Considering typical values of the coupling parameters in the Standard Model, it seems reasonable to assume that this ratio is indeed extremely small. Hence, we identify gravity with the slow sector while the matter field is considered to be the fast subsystem. 

The space adiabatic scheme requires a Hamiltonian formulation of the problem. We define the conjugate momenta of $a$ and $\cfv$ as, $p_a := \pp\frac{\partial L}{\partial \dot{a}}$ and $\cfm := \frac{\partial L}{\partial \dot{\phi}_{\scaleto{0}{3pt}}}$, where $L$ is the Lagrange function associated with the action $S$. The Poisson brackets of the canonical variables evaluate to $\lbrace a, p_a \rbrace = \pp$, and $\lbrace \phi_{0}, \mu_{0} \rbrace = 1$. The Legendre transformation generates the Hamilton constraint,
\begin{equation}
C(a,p_a,\phi_0,\pi_0) :=  -\frac{1}{12}\frac{p_a^2}{a} + \frac{\Lambda}{\lambda\,\kappa}a^3
+ \frac{\cfm^2}{2 a^3} +\frac{1}{2\lambda^2} m^2 a^3 \cfv^2,
\end{equation}
where for notational reasons, we divided the whole constraint by a constant factor $\lambda$. For simplifying the analysis by means of SAPT in the following, we switch to triad--like canonical variables,
\begin{equation}
\csv := \pm \sqrt{a^3}, ~~~~\csm :=\frac{2}{3} \frac{p_a}{\sqrt{a}},
\end{equation}
which is a double cover of the original phase space and we do not restrict to any of the branches of $b$. In order to keep the notation as simple as possible, we introduce the following parameters and functions,
\begin{equation}
\cmg := \frac{8}{3}, ~~~~~~~~ \ok^2 := \frac{3\,\Lambda}{4\,\lambda\,\kappa},
~~~~~~~~ \ml := b^2, ~~~~~~~~ \olam^2 := \frac{m^2}{\lambda^2}.
\end{equation}
These definitions and the new canonical variables give for the Hamilton constraint,
\begin{equation}
C(\csv,\csm,\cfv,\cfm) = - \frac{\csm^2}{2\,\cmg} + \frac{1}{2} \cmg \ok^2 \csv^2
+ \frac{\cfm^2}{2 \ml(\csv)} + \frac{1}{2} \ml(\csv) \olam^2 \cfv^2.
\end{equation}
We quantize the system and start by considering the scalar field subsystem using bold operator symbols. The state space is $\Hf:=L^2(\mathbb{R},\mathrm{d}\cfv)$, and the scalar field operator and its conjugate momentum satisfy the canonical commutation relation, $[\qcfv, \qcfm]_{\Lf} = i\,\Uf{\Lf}$. Similarly, the state space of the geometrical subsystem is $\Hs :=L^2(\mathbb{R},\mathrm{d}\csv)$. The quantum operators wear hats and the canonical commutation relation for the geometrical variable and its conjugate momentum are, $[ \qs{\csv},\qs{\csm}]_{\Ls} = i\,\pp\,\Us{\Ls}$. The quantum theory of the coupled system has the tensor product Hilbert space, $\mathcal{H} =\Hs \otimes \Hf$. The constraint operator on $\mathcal{H}$ is given by,
\begin{equation} \label{eq:Cosmological Model Quantum Constraint}
\qs{\qf{C}} = \left(- \frac{\qs{\csm}^2}{2 \cmg} + \frac{1}{2} \cmg\, \ok^2\, \qs{\csv}^2 \right) \otimes \Uf{\Lf} +  \frac{1}{2\,\ml(\qs{\csv})} \otimes \qf{\mu}_{0}^2 + \frac{1}{2} \ml(\qs{\csv})\, \olam^2 \otimes \qf{\phi}_{0}^2.
\end{equation}
We check the conditions (C1) -- (C4) for SAPT. (C1) holds without further ado since the cosmological Hilbert space $\Hs \otimes \Hf$ has the required tensor product structure, and $\Hs$ is an $L^2$--space and $\Hf$ is separable. We represent the quantum constraint as a symbol function $\qf{C}(\rho,b)$ with values in the linear operators on the Klein--Gordon Hilbert space $\Hf$ by formally quantizing the Klein--Gordon subsystem only, 
\begin{equation} \label{eq:FullHamiltonSymbol}
\qf{C}(\csv,\csm) =  \left(- \frac{\csm^2}{2 \cmg} + \frac{1}{2} \cmg \ok^2 \csv^2 \right) \Uf{\Lf}
+ \frac{\qcfv^2}{2\,\ml(\csv)}  + \frac{1}{2} \ml(\csv) \olam^2\, \qcfm^2.
\end{equation}
$\qf{C}(\csv,\csm)$ is an unbounded operator on $\Hf$ for every $(\csv,\csm) \in \mathbb{R}^2$. In particular, the operator corresponds to the Hamiltonian of a quantum harmonic oscillator with constant frequency $\olam$, $\csv$--dependent mass $\ml(\csv)$ and an off--set energy. As such, the symbol has for fixed finite $(\csv,\csm)$ an energy spectrum which is bounded from below but not from above. Besides, $\qf{C}(\csv,\csm)$ is an unbounded function with respect to both, $\csv$ and $\csm$. According to SAPT, the constraint symbol must however belong to one of the symbol classes $S^m_{\rho}(\mathcal{B}(\Hf))$ and should therefore have values in the space of \emph{bounded} operators on $\Hf$, be a bounded function with respect to $\csv$ and grow maximally polynomially in $\csm$. By means of the standard quantum oscillator eigensolutions $\xi_n \in \Hf$, $n\in\mathbb{N}$ with a $\csv$--dependent mass, the correspondent eigenvalue equation has the form,
\begin{equation}
\qf{C}(\csv,\csm)\,\xi_n(\csv) = E_n(\csv,\csm)\,\xi_n(\csv),~~~~ E_n(\csv,\csm) =  - \frac{\csm^2}{2 \cmg} + \frac{1}{2} \cmg \ok^2 \csv^2 + \olam\left(n+ \frac{1}{2} \right)
\end{equation}
We emphasize that the $\csv$--dependence of the states is purely parametric which allows to define $\csv$--dependent projection operators on $\Hf$, 
\begin{equation}
\qf{\pi}_n(\csv) := \xi_n(\csv)\,\langle \xi_n(\csv),\cdot\,\rangle_{\Hf},
\end{equation}
by means of which the Hamilton symbol constraint has the spectral representation,
\begin{equation} \label{eq:Cosmo Toy Constraint Spectral}
\qf{C}(\csv,\csm) = \sum_{n\geq0} E_n(\csv,\csm)\,\qf{\pi}_n(\csv).
\end{equation}
In order to respect the conditions for the application of SAPT, it is possible to define an auxiliary Hamilton symbol $\qf{C}_{\te{aux}}(b,\rho)$ which has values in the bounded operators, is locally a bounded function with respect to $b$, and which preserves the local structure of the symbol function $\qf{C}(b,\rho)$, \parencite{PST03,Stottmeister2015}. Since the perturbation scheme is applicable without referring to this auxiliary symbol (if convergence and uniformity of the series expansion do not play a role for the time being), we continue working with the original symbol \eqref{eq:Cosmo Toy Constraint Spectral} to illustrate the scheme. Most importantly, the gap condition (C3) is satisfied since the energy functions $E_n(b,\rho)$ are gapped functions on the gravitational phase space. Finally, we formally choose one of the fast energy bands with quantum number $\nu \in \mathbb{N}$ to proceed with the space adiabatic scheme.
\subsubsection*{Application of Space Adiabatic Perturbation Theory}
We start with the perturbative construction of the Moyal projector symbol $\qf{\pi}$ up to first order in $\pp$. In fact, this will be sufficient to define the effective Hamilton constraint up to \emph{second} order. With the ansatz, $\qf{\pi}_{(1)} = \qf{\pi}_0 + \pp \qf{\pi}_1$, and the natural choice for the base clause,
\begin{equation}
\qf{\pi}_{0} := \xi_\nu(\csv)\,\langle \xi_\nu(\csv),\cdot\,\rangle_{\Hf},
\end{equation}
we construct the symbol function $\qf{\pi}_{(1)}(\csv,\csm)$ following the construction steps (S1). The first condition (S1--1), $\qf{\pi} \star_\pp \qf{\pi} = \qf{\pi}$, yields that the diagonal contribution to $\qf{\pi}_{1}$ vanishes because $\qf{\pi}_{0}(\csv)$ depends solely on $\csv$. Regarding the third condition (S1--3), $\qf{C}_{0} \star_\pp \qf{\pi} - \qf{\pi} \star_\pp \qf{C}_{0} = 0$, it is straightforward to derive, \parencite{Teu03,NeuserSchanderThiemannII}, that,
\begin{align}
  \qf{\pi}_{1} = -\frac{i}{2} \,\qf{\pi}_{0}\cdot \lbrace \qf{\pi}_{0} , \qf{C}_{0} + E_\nu\, \qf{1}_{\te{f}} \rbrace_{\Ls} \cdot (\qf{C}_{0}^{\perp}- E_\nu\,\qf{1}_{\te{f}})^{-1}\cdot \qf{\pi}_{0}^\perp - \frac{i}{2} \,(\qf{C}_{0}^{\perp}- E_\nu\,\qf{1}_{\te{f}})^{-1} \cdot \qf{\pi}_{0}^\perp \cdot  \lbrace \qf{C}_{0} + E_{\nu}\,\qf{1}_{\te{f}},\qf{\pi}_{0}   \rbrace_{\Ls} \cdot \qf{\pi}_{0},
\end{align}
as a determining equation for $\qf{\pi}_{1}$, where we defined, $\qf{C}_{0}^{\perp} = \qf{C}_{0}\cdot\qf{\pi}_{0}^\perp$, and $\qf{\pi}_0^\perp := \qf{1}_{\te{f}}- \qf{\pi}_0$. To evaluate the partial derivative $\partial_\csv \qf{\pi}_{0}$ in this equation, we need to evaluate the derivative of the states $\xi_n(b) \in \Hf$ as well as the derivatives of the canonically defined creation and annihilation operators $\qf{a}^\ast(b) \in \mathcal{L}(\Hf)$ and $\qf{a}(b) \in \mathcal{L}(\Hf)$. Therefore, recall that the initial eigenvalue problem admits the oscillator solutions $\xi_n(b)$ . Accordingly, the creation operator $\qf{a}^\ast(b)$  can be written in terms of the canonical pair $(\qf{\phi}_{0},\qf{\mu}_{0})$ as,
\begin{equation}
\qf{a}^\ast(\csv) = \sqrt{\frac{\ml(\csv) \olam}{2}} \left( \qcfv
- \frac{i}{\ml(\csv) \olam} \qcfm \right),
\end{equation}
The derivatives of the vacuum state $\xi_{0}(b)$ and the creation operator are given by,
\begin{equation}
  \frac{\partial \xi_{0}}{\partial \csv} := \sqrt{2} f(\csv)\,\xi_{2}(\csv), ~~~ \frac{\partial \qf{a}^\ast(\csv)}{\partial \csv} = -2 f(\csv)\, \qf{a}(\csv),
\end{equation}
where $f(\csv):= - (\partial_{\csv} \ml)/(4\ml) = - 1/(2\csv)$. We propose the definition of a covariant derivative, or more precisely, a gauge potential $\qf{\mathcal{A}}$, associated with the $\csv$--derivative of the fast oscillator states. Note that this is simply Berry's connection, \parencite{Berry1984}. Using the natural basis choice from above, its coefficients with respect to the $b$--direction on $\Gamma_{\te{s}}$ are given by,
\begin{equation}
  \frac{\partial \xi_n(\csv)}{\partial \csv} = \mathcal{A}_{\csv n}^{~~n-2}(\csv)\, \xi_{n-2}(\csv) + \mathcal{A}_{\csv n}^{~~n+2}(\csv)\, \xi_{n+2}(\csv),
\end{equation}
with, $\mathcal{A}_{bn}^{~k}(\csv) = -\sqrt{n(n-1)}\,f(\csv)\,\delta_n^{k+2} + \sqrt{(n+1)(n+2)}\,f(\csv)\,\delta_n^{k-2}$. All coefficients $\mathcal{A}_{\rho n}^{~~m}$ in the $\rho$--direction vanish because the fast eigenstates do not depend on $\rho$.  Only the coefficients that connect states differing by two excitations in the $b$--direction are non--vanishing. Since we have real--valued eigenstates, the connection coefficients are real--valued, too, such that the orthonormality relation between the fast states yields that, $\mathcal{A}_{\csv n}^{~~m} = - \mathcal{A}_{bm}^{~~n}$. The $b$--derivative of the projector symbol $\qf{\pi}_{0}$ follows from using Riesz' representation theorem and one can simply write,
\begin{equation}
\frac{\partial \qf{\pi}_{0}}{\partial \csv} = \mathcal{A}_{\csv \nu}^{~~m} \left( \xi_\nu \langle \xi_m,\cdot\,\rangle_{\te{f}} + \xi_m \langle \xi_\nu,\cdot\,\rangle_{\te{f}} \right),
\end{equation}
where $\nu$ is still a fixed quantum number while $m$ runs over all natural numbers.
To evaluate $\qf{\pi}_{1}$, the partial derivative, $\partial_{\csm}\,(\qf{C}_{0}+E_{\nu}\cdot\Uf{\Lf})$, is simply $(-2\csm/\cmg)\cdot \Uf{\Lf}$, because only the spectral functions $E_n(\csv,\csm)$ depend on $\csm$ while the states do not. The functional form of the energy functions also reduces $(\qf{C}_{0}^{\perp}-E_{\nu}\cdot\Uf{\Lf})$ to a factor $\pm (2\olam)^{-1}$, and consequently,
\begin{align}
  \qf{\pi}_{1}\! = \!-\frac{i \csm}{2 \cmg \olam} \!\left( \mathcal{A}_{\csv \nu}^{~~\nu-2} \left( \xi_\nu \langle \xi_{\nu-2},\cdot\, \rangle_{\Lf}\! - \xi_{\nu-2} \langle \xi_\nu,\cdot\, \rangle_{\Lf} \right) + \mathcal{A}_{\csv\nu}^{~~\nu+2} \left(\xi_{\nu+2} \langle \xi_\nu,\cdot\, \rangle_{\Lf} \!- \xi_\nu \langle \xi_{\nu+2},\cdot\, \rangle_{\Lf} \right) \right).
\end{align}
One can easily check that $\qf{\pi}_{(1)}$ satisfies all three conditions subsumed under (S1) up to first order in $\pp$, i.e., that it is a projector and commutes with the full Hamilton symbol up to errors of order $\pp^2$. We see that the improved projection symbol mixes adjacent eigenstates of the fast system, and going to higher orders in the perturbative scheme more and more states will be included.

The next step of SAPT consists in constructing the unitary symbol $\qf{u}_1$ which maps the dynamical subspace related to $\qf{\pi}_{(1)}$, to a suitable reference subspace $\mathcal{K}_{\Lf} \subset \Hf$. It is convenient to choose one point $(b_{0},\rho_{0}) \in \Gamma_{\Ls}$ and define the reference projection as,
\begin{equation}
\qf{\pi}_{\te{R}} := \xi_\nu(b_{0}) \left\langle  \xi_\nu(b_{0}), \cdot \,\right\rangle_{\Lf} =: \zeta_\nu \langle \zeta_\nu, \cdot \,\rangle_{\Lf},
\end{equation}
where $\zeta_\nu \in \Hf$ does not depend on the gravitational phase space variables. In a similar fashion, one can define the complete basis $\zeta_n := \xi_n(b_0)$, $n\in \mathbb{N}$ at the point $b_0$. A natural choice for the unitary operator in line with conditions (S2) at zeroth order is simply,
\begin{equation}
\qf{u}_0(b) = \sum_{n \geq 0} \zeta_n \left\langle \xi_n(b), \cdot\, \right\rangle_{\Lf}.
\end{equation}
In order to construct $\qf{u}_1$, the scheme splits the symbol into a hermitian and an antihermitian part. The hermitian part is determined by equations (S2--1) and (S2--2), namely by requiring that $\qf{u} \star \qf{u}^\ast = \qf{1}_{\te{f}}$ holds up to first order in $\pp$. Since $\qf{u}_0$ only depends on the configuration variable $b$, the hermitian part vanishes trivially. The antihermitian part is determined by restricting equation (S2--3), i.e., $\qf{u} \star \qf{\pi}\star\qf{u}^\ast = \qf{\pi}_{\te{R}}$, to the first order. It yields for $\qf{u}_{1}$, \parencite{NeuserSchanderThiemannII},
\begin{align}
  \qf{u}_{1} &= [\qf{\pi}_{\te{R}}, \qf{u}_{0}\cdot \qf{\pi}_{1}^{\te{OD}}\cdot \qf{u}_{0}^{\ast}]_{\text{f}}\cdot \qf{u}_{0} \\
  &= \frac{i \csm}{2\, \cmg\, \olam} \!\left[ \mathcal{A}_{\csv \nu}^{~~\nu-2} \left( \zeta_\nu \langle \xi_{\nu-2},\cdot \rangle_{\Lf} + \zeta_{\nu-2} \langle \xi_\nu,\cdot\, \rangle_{\Lf} \right) - \!\mathcal{A}_{\csv\nu}^{~~\nu+2} \!\left( \zeta_{\nu+2} \langle \xi_\nu,\cdot\, \rangle_{\Lf} + \zeta_\nu \langle \xi_{\nu+2},\cdot\, \rangle_{\Lf} \right) \right]. \nonumber
\end{align}
Eventually, we are ready to compute the effective Hamiltonian symbol up to second order in the perturbations, and which we restrict to the selected reference space, i.e., we compute, $\qf{C}_{\text{eff},(2),\te{R}}(\csv,\csm) := \qf{\pi}_{\te{R}} \cdot \qf{C}_{\text{eff},(2)} (\csv,\csm) \cdot \qf{\pi}_{\te{R}}$. The zeroth order contribution of this symbol is given according to condition (S3) by,
\begin{equation} \label{eq:ZeroHeff}
\qf{C}_{\text{eff},0,\te{R}}(\csv,\csm)
= \left( -\frac{\csm^2}{2 \cmg} + \frac{1}{2} \cmg \ok^2 \csv^2
+ \olam  \left( \nu + \frac{1}{2} \right) \right) \qf{\pi}_{\te{R}}.
\end{equation}
Thus, the effective constraint symbol for the gravitational degrees of freedom includes the bare gravitational constraint symbol plus an off--set energy which stems from the energy band associated with the quantum number $\nu$ of the Klein--Gordon system. This result corresponds to the Born--Oppenheimer approximation. The first order contribution of the effective constraint symbol, $\qf{C}_{\text{eff},1}(\csv,\csm)$ contains only off-diagonal terms, such that $\qf{C}_{\text{eff},1,\te{R}}(\csv,\csm)$ vanishes identically,
\begin{equation}
\qf{C}_{\text{eff},1,\te{R}}(\csv,\csm) = \frac{i}{2} \left\lbrace \qf{\pi}_{\te{R}} \cdot  \qf{u}_{0}, \qf{C}_{0}
+ E_{\nu}\,\Uf{\Lf} \right\rbrace_{\Ls}\cdot \qf{u}_{0}^{\ast}\cdot \qf{\pi}_{\te{R}} =0.
\end{equation}
The same reasoning applies to the computation of the second order contribution $\qf{C}_{\text{eff},2,\te{R}}(\csv,\csm)$ giving,
\begin{align} \label{second}
\qf{C}_{\text{eff},2,\te{R}} &= \frac{i}{2} \lbrace \qf{\pi}_{\te{R}} \cdot \qf{u}_{1}, \qf{C}_{0} + E_\nu \,\qf{1}_{\te{f}} \rbrace_{\te{s}} \cdot \qf{u}_{0} \cdot \qf{\pi}_{\te{R}}\\
&= \left[\frac{\partial E_n}{\partial \csm} \right]^2 \left[ \frac{(\mathcal{A}_{\csv \nu}^{~~\nu-2})^2}{E_\nu - E_{\nu-2}} + \frac{(\mathcal{A}_{\csv \nu}^{~~\nu+2})^2}{E_\nu - E_{\nu+2}} \right]\qf{\pi}_{\te{R}} + \frac{1}{2} \frac{\partial^2 E_n}{\partial \csm^2} \left[ (\mathcal{A}_{\csv \nu}^{~~\nu-2})^2+(\mathcal{A}_{\csv \nu}^{~~\nu+2})^2\right] \qf{\pi}_{\te{R}}. \nonumber
\end{align}
Finally, inserting the explicit results for the energy functions and the connection coefficients yields,
\begin{equation} \label{eq:Cosmo Toy model solution}
\qf{C}_{\text{eff},2,\te{R}}(\csv,\csm) = -\frac{1}{2\,\cmg} \left(  \frac{\csm^2}{\cmg \olam \csv^2} \left(\nu + \frac{1}{2} \right) + \frac{1}{2\csv^2}  \left(\nu^2 + \nu+1 \right)\right) \qf{\pi}_{\te{R}}.
\end{equation}
This proves our statement that besides the trivial Born--Oppenheimer approximation, further backreaction effects arise for the gravitational subsystem. It is now easy to evaluate the action of this symbol on some generic tensor product wave function in $\mathcal{H} = \Hs \otimes \Hf$, since the Klein--Gordon tensor factor does not depend on the gravitational degrees of freedom anymore. The problem reduces to a quantum constraint equation on $\Hs$ only, and can be studied for each $\nu$ of interest. Nevertheless, we point out that finding states that are annihilated by the constraint operator $\qs{\qf{C}}_{\te{eff},2,\te{R}}$ is not a trivial task as it depends non--polynomially on $b$. Further details can be found in \parencite{NeuserSchanderThiemannII}. Finally, the question is how the solutions of \eqref{eq:Cosmo Toy model solution} relate to the solutions of the original problem on $\mathcal{H}$. In fact, one needs to rotate the solutions of $\qs{\qf{C}}_{\te{eff},2,\te{R}}$ back to the original Hilbert space using the quantization of the Moyal unitary. It turns out that the resulting solutions are also approximate (orthogonal) solutions to the full Hamilton constraint at the respective perturbative order, \parencite{Teu03,SchanderThiemannI}.

\subsection{Backreaction in Inhomogeneous Quantum Cosmology} 
The purely homogeneous model in the previous section can serve as a showcase for a more realistic inhomogeneous cosmological model. Here, we consider standard cosmological perturbation theory for a gravitational metric field $g$, a massive real scalar field $\Phi$ as the matter content, and a cosmological constant, $\Lambda >0$. After the split of the relevant degrees of freedom into a homogeneous and an inhomogeneous part, the aim will be to incorporate backreactions from the perturbative degrees of freedom onto the homogeneous and isotropic background degrees of freedom. This section relies on \parencite{SchanderThiemannIV} to which we refer for more details.

As before, the model rests on a four--dimensional globally hyperbolic space time manifold $\mathcal{M}$ that admits the time space split $\mathcal{M} \cong \mathbb{R} \times \sigma$. The metric has Lorentzian signature $(-,+,+,+)$, and the spatial hypersurfaces $\sigma$ are compact and flat three--tori $\sigma \cong \mathbb{T}^3$ with side lenghts $l \equiv 1$. The global time parameter $t \in \mathbb{R}$ labels the spatial Cauchy surfaces $\Sigma_t$. $n^{\mu}$ is the unit normal vector field to these hypersurfaces, $N$ and $N_{\mu}$ the (standard) lapse and shift functions which parametrize the normal and the tangential part of the foliation. The task of specifying constraints or equations of motion for the metric field $g$, translates into defining a Cauchy initial value problem for the spatial metric $h_{\mu \nu} = g_{\mu \nu} + n_{\mu} n_{\nu}$ induced by $g$. The extrinsic curvature associated with the time derivative of $h$ is given by $K_{\mu \nu} = h_{\mu}^{\rho} h_{\nu}^{\lambda} \nabla_{\rho} n_{\lambda}$. $\nabla$ is the unique, torsion--free covariant derivative associated with the metric $g$. After pulling back the tensor fields to $\mathbb{R}\times \mathbb{T}^3$ and denoting spatial indices on the spatial hypersurfaces with lower case latin symbols, $a,b,c,..\in \left\lbrace1,2,3 \right\rbrace$, the Lagrange density is expressed by the sum of the Einstein--Hilbert Lagrange density $\mathcal{L}_{\te{EH}}$ and the scalar field Lagrange density $\mathcal{L}_\Phi$, with,
\begin{equation} \label{eq:LEH}
\mathcal{L}_{\te{EH}} = \frac{1}{2 \kappa} \sqrt{|h|} N \left( R^{(3)} + K_{ab} K^{ab} - (K_a^a)^2 - 2 \Lambda \right),
\end{equation}
\begin{equation} \label{eq:LPhi}
\mathcal{L}_\Phi =  \frac{1}{2\lambda} \sqrt{|h|} N \left( - \frac{1}{N^2} \dot{\Phi}^2 + 2 \frac{N^a}{N^2} \dot{\Phi} \partial_a \Phi + \left( h^{ab} - \frac{N^a N^b}{N^2} \right) \partial_a \Phi \partial_b \Phi + m^2 \Phi^2 \right).
\end{equation}
where again, $\lambda$ is the coupling constant of the scalar field, $m$ is the mass parameter of the scalar field, and $R^{(3)}$ is the curvature scalar associated with the three--metric $h$ and its Levi--Civita covariant derivative, $D$. The only degrees of freedom of the spatially homogeneous and isotropic sector are the zeroth order lapse function $N_0(t)$ and the scale factor $a(t)$, associated with the zeroth order spatial metric, $h^0(t,x) := a^2(t)\tilde{h}^0(x)$ with $\tilde{h}^0(x)$ being the time--independent metric on the spatial hypersurfaces. A Hamiltonian analysis shows that the lapse function is a Lagrange multiplier with no dynamical features, affirming the arbitrariness of the hypersurface foliation. 

We introduce perturbations of the homogeneous degrees of freedom using a decomposition into scalar, vector and tensor parts, \parencite{HalliwellHawking1987}. Since we make use of their results in a later stage, we will stick to the definition of perturbations used by \textcite{CMM15,MartinezOlmedo2016}, 
\begin{align}
N(t,x) &=: N_0(t) + a^3(t)\, g(t,x) \label{eq:Lapse} \\
N_a(t,x) &=: a^2(t)\,D_a\, k(t,x) + a^2(t)\,\epsilon_a^{\; b c} D_b\,k_c(t,x) \label{eq:Shift} \\
h_{ab}(t,x) &=:  a^2(t) \left[\left(1+2\,\alpha(t,x) \right) \!\tilde{h}^0_{ab}(x) + 6 \Big( D_a D_b - \frac{1}{3} \tilde{h}^0_{ab}(x) D_c D^c \Big) \beta(t, x) + 2 \sqrt{6} \, t_{ab}(t, x) \right. \nonumber \\
 &~~~~~~~~~~~~~~~~~~ \left. + \,4 \sqrt{3} D_{(a} v_{b)}(t,x) \right],\label{eq:Perturbation Metric} \\
\Phi(t,x) &=: \phi(t) + \varphi(t,x). \label{eq:phi}
\end{align}
where we introduced the perturbative scalar fields $(g, k, \alpha, \beta, \varphi )$, the vector fields $v_a$ and $k_a$, and the tensor field perturbations $t_{a b}$. For notational reasons, we introduce the fields $\check{k}:=\Delta k$ and $\check{k}_a:= \epsilon_a^{\;bc} D_b k_c$ as new degrees of freedom associated with the shift. 

We perform a Legendre transformation to obtain the Hamilton constraint. We insert the perturbed variables from equations \eqref{eq:Lapse} -- \eqref{eq:phi} into the Lagrange density \eqref{eq:LEH}, \eqref{eq:LPhi}, and expand the Lagrangian and the action functional $S$ up to second order in the perturbations. As the three--torus does not have a boundary, total divergences vanish in the computations. The resulting action does neither depend on the velocities of the lapse variables $N_0$ and $g$, nor on the velocities of the shift variables $\check{k}$ and $\check{k}_a$. These Lagrange multipliers will hence be associated with primary constraint equations in the Hamiltonian formalism. In the lines of \parencite{CMM15}, we define the conjugate momenta $(P_a, P_{\phi})$ for the homogeneous and isotropic degrees of freedom $(a,\phi)$ using the Lagrange function $L=\int \mathrm{d} x\,\mathcal{L}$,
\begin{equation}
  P_a := \frac{\partial L}{\partial \dot{a}} = - \frac{6}{\kappa N} a \dot{a},~~~~~~~~~ P_\phi := \frac{\partial L}{\partial \dot{\phi}} = \frac{a^3}{\lambda N} \dot{\phi}.
\end{equation}
We denote the corresponding phase space by $\Gamma_{\!\text{hom}} = \Gamma_{\te{s}}$. The perturbation fields $(\alpha, \beta, \varphi, \vartheta_a, t_{ab})$ together with their conjugate momenta $(\pi_\alpha, \pi_\beta, \pi_\varphi, \pi^a_v, \pi^{ab}_t)$ span the perturbative phase space $\Gamma_{\!\text{pert}} = \Gamma_{\te{f}}$. The momenta are defined as,
\begin{equation}
  \pi_\chi := \frac{\partial \mathcal{L}}{\partial \dot{\chi}},
\end{equation}
for any field $\chi \in \lbrace \alpha, \beta, \varphi, v_a, t_{ab} \rbrace$. $N_0$, $g$, $\check{k}$ and $\check{k}_a$ induce the lapse and shift primary constraints $\Pi_0^{N_0}$, $\Pi_1^{g}$, $\Pi_1^{\check{k}}$ and $\Pi_1^{\check{k}_a,b}$. The Hamiltonian density is eventually given by,
\begin{align}
\mathcal{H} =&~ N_0 \left[ \mathcal{H}_0 + \mathcal{H}_2^{\text{s}} + \mathcal{H}_2^v + \mathcal{H}_2^{t} \right] + g \cdot \mathcal{H}_1^{g} + \check{k}_a \cdot \mathcal{H}_1^{\check{k}_d,a} + \check{k} \cdot \mathcal{H}_1^{\check{k}} \nonumber \\
& + \lambda_{N_0}\cdot \Pi_0^{N_0} + \lambda_{g} \cdot \Pi_1^{g} + \lambda_{\check{k}}\cdot\Pi_1^{\check{k}} + \lambda_{\check{k}_a,b} \cdot \Pi_1^{\check{k}_a,b}. \label{eq:H original}
\end{align}
$\mathcal{H}_0$ denotes the Hamiltonian contribution associated with the completely homogeneous and isotropic model. $\mathcal{H}_2^{\text{s}}$, $\mathcal{H}_2^v$ and $\mathcal{H}_2^t$ are of second order in the perturbations and contain only scalar, vector and tensor variables respectively. $\mathcal{H}_1^g$, $\mathcal{H}_1^{\check{k}_d,a}$ and $\mathcal{H}_1^{\check{k}}$ represent first order contributions which factorize with the respective lapse and shift variables. The second line lists the primary constraints associated with lapse and shift and their Lagrange multipliers $\lambda_{N_0}$, $\lambda_g$, $\lambda_{\check{k}}$ and  $\lambda_{\check{k}_a,b}$. The system is completely constrained and we thus perform a Dirac analysis to extract the relevant physics. 

Therefore, we identify a suitable set of free variables -- in fact, the Dirac analysis will then become a trivial task. We start by noting that the perturbation variables that we introduced are not all gauge--invariant. In the scalar sector, let us employ the gauge--invariant Mukhanov-Sasaki variable $\vartheta$, \parencite{Mukhanov1988,Mukhanov2005},
\begin{equation}
  \vartheta := a\,\varphi + \frac{6 \lambda P_\phi}{\kappa P_a} (\alpha - \Delta \beta)
\end{equation}
Note that this transformation for the perturbative fields also depends on the homogeneous degrees of freedom. While the original perturbation variables had canonical momenta properly defined by the Legendre transform, the new perturbation variables will break the canonical structure as they non--trivially depend on the homogeneous degrees of freedom. In order to preserve the canonical structure of the system, it is mandatory to find a suitable transformation for the homogeneous and isotropic variables, too. This appears to be a cumbersome mission. \textcite{CMM15} have however shown that it is possible to find a transformation for the homogeneous and isotropic degrees of freedom which preserves the canonical structure of the system up to second order in the cosmological scalar perturbations. The same has been done by \textcite{MartinezOlmedo2016} for the tensor degrees of freedom.

These transformations and the resulting almost--canonical variables in the homogeneous and the perturbative sector lead to a redefinition of the Hamilton constraint. Fortunately, these transformations make the Dirac constraint analysis almost trivial. The secondary constraints that one obtains from requiring the conservation of the primary constraints under the evolution of the Hamiltonian, generate only one single, non--trivial constraint, namely,
\begin{equation}
\mathcal{C} := \mathcal{H}_0 + \tilde{\mathcal{H}}_2^s +\Check{\mathcal{H}}_2^t =0, \label{eq:Hamilton Constraint}
\end{equation}
where $\mathcal{H}_0$ now depends on the transformed homogeneous variables, and $\tilde{\mathcal{H}}_2^s$ and $\Check{\mathcal{H}}_2^t$ are second order constraints that depend on the Mukhanov--Sasaki and the tensor perturbations respectively, as well as on the transformed homogeneous variables. Before we give the corresponding expressions, let us perform the typical $\pp$--rescalings of the momenta that will make the space adiabatic scheme work at the technical level. We will simply denote the transformed variables by the original ones, and write,
\begin{equation}
p_a := \pp^2 \,P_a,~~~~~~~~~ p_\phi := \pp\,P_\phi.
\end{equation}
Besides, we perform a canonical rescaling of the inhomogeneous Mukhanov--Sasaki and tensor perturbations, and multiply the constraint by a global $\pp^2$--factor. Then, the total Hamilton constraint, $C = H_0 + \tilde{H}_2^s + \Check{H}_2^t =0$, is given by, \parencite{SchanderThiemannIV},
\begin{align} \label{eq:Cosmological Perturbations H_0 classical}
H_0 &= - \frac{\Gm^2}{12 \Gv} + \frac{\KGm^2}{2 \Gv^3} + \frac{1}{2} \varepsilon^2 m^2 \Gv^3 \KGv^2 + \Lambda \Gv^3, \\
\tilde{H}_2^s &= \frac{1}{2 \Gv} \int_{\mathbb{T}^3} \mathrm{d} x\,\left[ \frac{\MSm^2}{\sqrt{\tilde{h}^0}}  + \MSv\, \pp^4 \left[ - \sqrt{\tilde{h}^0} \Delta + \mMS^2 \right] \MSv  \right],  \label{eq:Cosmological Perturbations H_2 MS classical}\\
\Check{H}_2^t &= \frac{1}{2\Gv} \int_{\mathbb{T}^3} \mathrm{d} x\, \left[\frac{\pi_t^{ab} \pi_{t,ab} }{6\sqrt{\tilde{h}^0}}  + t^{ab} \pp^4 \left[ - 3 \sqrt{\tilde{h}^0} \Delta + (\pp\,\mT)^2  \right] t_{ab} \!\right], \label{eq:Cosmological Perturbations H_2 Tensor classical}
\end{align}
with the effective Mukhanov--Sasaki and tensor masses depending on the homogeneous variables,
\begin{align} \label{eq:Cosmological Perturbations MS mass final}
&\mMS^2 = \left(- \frac{\Gm^2}{18 \Gv^2} + \frac{7 \KGm^2}{2 \Gv^4} -  12 \pp \sqrt{\tilde{h}^0}^{\,2} m^2 \frac{\Gv \KGv \KGm}{\Gm} - 18\,\frac{\KGm^4}{\Gv^6 \Gm^2} + \sqrt{\tilde{h}^0}^{\,2} m^2  \Gv^2 \right) \frac{1}{\sqrt{\tilde{h}^0}}, \\
&(\pp\,\mT)^2 = \left(\frac{\Gm^2}{6 \Gv^2} - 3 \pp^2 \sqrt{\tilde{h}^0}^{\,2}\,m^2 \Gv^2 \KGv^2 - 6 \sqrt{\tilde{h}^0}^{\,2} \Lambda \Gv^2 \right) \frac{1}{\sqrt{\tilde{h}^0}}. \label{eq:Cosmological Perturbations Tensor mass final}
\end{align}
Now, SAPT requires to quantize the cosmological perturbations. The form of $\tilde{H}_2^s$ and $\Check{H}^t_2$ suggests to consider a standard Fock quantization for the Mukhanov--Sasaki and the tensor fields. The quantized fields should satisfy the commutation relations,
\begin{equation}
\left[ g(\qf{\MSv}), \qf{\pi}_\MSv(f) \right]_{\te{MS}} =i \int_{\mathbb{T}^3}\!\mathrm{d}x\,g(x) f(x) \,\Uf{\te{MS}}, ~~~\left[ G(\qf{t}), \qf{\pi}_t(F)\right]_{\te{T}} = i\int_{\mathbb{T}^3} \!\mathrm{d}x\,G^{ab}(x)F_{ab}(x) \,\Uf{\te{T}}, \nonumber
\end{equation}
for suitable test functions $f$, $g$, $G^{ab}$ and $F_{ab}$ on the three--torus, where $\Uf{\te{MS}}$ and $\Uf{\te{T}}$ denote the unities of the quantum algebras. As a basis for the one--particle Hilbert space $L^2(\mathbb{T}^3,\mathrm{d}x)$ it is convenient to choose the plane waves $f_{\vec{k}}(\vec{x}) = \exp (i \vec{k} \vec{x})$, with $\vec{k} \in 2 \pi \mathbb{Z} \setminus 0$ being discrete. The total perturbative Hilbert space is the tensor product of the correspondent Fock spaces,
\begin{equation}
\Hf = \mathcal{F}_{\text{s,MS}}(L^2(\mathbb{T}^3,\mathrm{d}x)) \bigotimes_{\tau =\lbrace +,- \rbrace} \mathcal{F}_{\te{s},\te{T},\tau}(L^2(\mathbb{T}^3,\mathrm{d}x)),
\end{equation}
where the index ``s'' refers to symmetric, and we introduced the label $\tau$ which stands for the only two physical degrees of freedom associated with the tensor perturbations (namely their polarizations). The form of the second order contributions to the Hamilton constraint suggest to define the one--particle frequency operators for the Mukhanov--Sasaki and the tensor systems,
\begin{equation} \label{eq:Def Frequencies}
\omega_{\te{MS}}:= \pp^2 \sqrt{- \Delta + \mMS^2}, ~~~~ \omega_{\te{T}}:= \pp^2 \sqrt{-18 \Delta + 6\,(\pp\, \mT)^2}.
\end{equation}
Note that both operators depend on the homogeneous degrees of freedom as they contain the mass functions $m_{\text{MS}}(a,p_a,\phi,p_\phi)$ and $m_{\text{T}}(a,p_a,\phi)$. This implies that the annihilation and creation operators $\qf{a}(f_{\vec{k}}) =: \qf{a}_{\vec{k}}$, $\qf{a}^\ast(f_{\vec{k}})=:\qf{a}_{\vec{k}}^\ast$ of the Mukhanov--Sasaki system and the annilation and creation operators of the tensor perturbations $\qf{b}_\pm(f_{\vec{k}}) = \qf{b}_{\vec{k},\pm}$, $\qf{b}_\pm^\ast(f_{\vec{k}}) = \qf{b}_{\vec{k},\pm}^\ast$, defined in the standard way, depend on the homogeneous degrees of freedom, and so do the natural Fock basis states, \parencite{SchanderThiemannIV}. In contrast to the cosmological toy model, they depend on the whole set of phase space variables, which represent non--commuting operators in the quantum theory, and which makes the application of SAPT mandatory. The Born--Oppenheimer approach would fail in the given case.

It is most convenient to express the quantum constraint symbol $\qf{C}(a,p_a,\phi, p_\phi)$ in terms of normal--ordered annihilation and creation operators,
\begin{align}
\qf{C} =\left(- \frac{\Gm^2}{12 \Gv} + \frac{\KGm^2}{2 \Gv^3} + \frac{1}{2}\,\varepsilon^2\,\mPhi^2 \Gv^3\,\KGv^2 + \Lambda\,\Gv^3 \!\right)\qf{1}_{\text{f}} + \frac{1}{\Gv} \sum_{\vec{k} \in \mathbb{k}} \omega_{\text{MS},\vec{k}}\; \adMS{\vec{k}}\,\aMS{\vec{k}} +  \frac{1}{6 \Gv} \sum_{\vec{K} \in \mathbb{K}} \omega_{\text{T},\vec{k}} \;\bdT{\vec{K}}{}\,\bT{\vec{K}}{}, 
\end{align}
where we identify the first contribution as the usual FLRW Hamilton constraint, which we will denote by $E_{\text{hom}}(\Gv,\Gm,\KGv,\KGm)$. The label $\vec{K} := (\vec{k},\tau) \in \mathbb{K}$ specifies the mode and the polarization of the tensor perturbations. $\qf{C}$ admits a discrete spectrum for any point $(a,p_a,\phi,p_\phi) \in \Gamma_{\text{s}}$ because the sums over the wave vectors in the Hamilton constraint are discrete and so is the spectrum of the number operators $\qf{a}_{\vec{k}}^\ast\, \qf{a}_{\vec{k}}$ and $\qf{b}_{\vec{K}}^\ast\, \qf{b}_{\vec{K}}$. Any Fock state $\xi_{(n)} \in \Hf$ with finite energy can be identified with a finite set of non--vanishing quantum numbers $(n) := \lbrace \dots, n_{\text{MS},\vec{k}_1}, n_{\text{MS},\vec{k}_2}, \dots, n_{\text{T},\tau,\vec{k}_1}, n_{\text{T},\tau,\vec{k}_2}, \dots  \rbrace$, where we distinguished between the quantum numbers of the Mukhanov--Sasaki and the tensor perturbations. We introduce degeneracy labels which take the possibility of degenerate eigenstates into account, and we denote them by $b = 1, \dots, d$ for the Mukhanov--Sasaki system and $b' = 1, \dots, d'$ for the graviton system. To shorten notation, we integrate the degeneracy labels in $\beta := \lbrace b,b' \rbrace$ and the degeneracy numbers in $\delta := \lbrace d, d' \rbrace$. We write for the set of homogeneous variables, $(q,p):= (\Gv,\Gm,\KGv,\KGm)$. The eigenvalue problem for any finite set of quantum numbers $(n)_\beta$ then has the form,
\begin{align}
&\qf{C}(q,p)\, \xi_{(n)_\beta}(q,p) = E_n (q,p) \,\xi_{(n)_\beta} (q,p),\\
&E_{n}(q,p) := E_{\text{hom}}(a,p_a,\phi,p_\phi) + \frac{1}{a} \sum_{\vec{k} \in \mathbb{k}} n_{\text{MS},\vec{k},b}\, \omega_{\text{MS},\vec{k}} + \frac{1}{6a} \sum_{\vec{K} \in \mathbb{K}} n_{\text{T},\vec{K},b'}\, \omega_{\text{T},\vec{K}}. \nonumber
\end{align}
Due to the discreteness of the eigenbasis, it is possible to define non--vanishing energy gaps between the eigenenergy bands of the perturbations, at least for local regions in phase space. In the following, we assume that the relevant energy bands admit such local gaps in the region of interest.

To examine the derivatives of the eigenstates with respect to the homogeneous variables, we need the derivative of the vacuum state and the annihilation operators since any excited state in the Hilbert space $\Hf$ can be constructed from the vacuum state $\Omega(q,p)$ by applying the desired number $(n_{\te{MS},\vec{k}},\;n_{\te{T},\vec{k},+}, n_{\te{T},\vec{k},-})$ of creation operators for every wave number $\vec{k}$. We formally choose one such eigenstate with quantum number(s) $(\nu)_\beta$ given by,
\begin{equation} \label{eq:MST excited states}
\xi_{(\nu)}(q,p) = \prod_{\vec{K} \in \mathbb{K}} \frac{(\adMS{\vec{k}})^{\nu_{\te{MS},\vec{k}}}}{\sqrt{\nu_{\te{MS},\vec{k}}!}} \frac{(\bdT{\vec{K}})^{\nu_{\te{T},\vec{K}}}}{\sqrt{\nu_{\te{T},\vec{K}}!}}\,\Omega(q,p)
\end{equation}
It is useful to write the explicit representation of the Mukhanov--Sasaki wave function and the tensor wave functions as a product,
\begin{equation}
\xi_{(\nu)} =: \xi_{(\nu_{\te{MS}})}^{\te{MS}} \cdot \prod_{\tau}  \xi_{(\nu_{\te{T}})}^{\te{T},\tau}.
\end{equation}
The derivatives of the annihilation operators with respect to $\lambda \in \lbrace q,p \rbrace$, are proportional to the correspondent creation operators, 
\begin{align}
\frac{\partial \qf{a}_{\vec{k}}(q,p)}{\partial \lambda} := f_{\lambda, \vec{k}}^{\te{MS}}(q,p)\,\qf{a}^\ast_{\vec{k}}(q,p),~~~~
\frac{\partial \qf{b}_{\vec{K}}(q,p)}{\partial \lambda} := f_{\lambda, \vec{K}}^{\te{T}}(q,p)\,\qf{b}^\ast_{\vec{K}}(q,p),
\end{align}
and the explicit expressions of the factors can be found in \parencite{SchanderThiemannIV}. The $\lambda$--derivative of the vacuum state is then given by,
\begin{equation}
\frac{\partial \Omega(q,p)}{\partial \lambda} = \sum_{\vec{k} \in \mathbb{k}} f_{\lambda,\vec{k}}^{\te{MS}}(q,p) (\qf{a}_{\vec{k}}^\ast\,\qf{a}_{\vec{k}}^\ast\, \Omega)(q,p) + \sum_{\vec{K} \in \mathbb{K}} f_{\lambda,\vec{k}}^{\te{T}}(q,p) (\qf{b}_{\vec{K}}^\ast\,\qf{b}_{\vec{K}}^\ast\, \Omega)(q,p).
\end{equation}
The $\lambda$--derivative of any excited state $\xi_{(n)}$ is thus uniquely defined by these results and can be expressed as an application of a connection $\qf{\mathcal{A}}_\lambda \in C^\infty(\Gamma_{\te{s}},\mathcal{L}(\Hf))$, on the global Hilbert bundle $H$, 
\begin{equation} \label{eq:Def Connection}
\frac{\partial \, \ef{(n)}(q,p)}{\partial \lambda} =: \qf{\mathcal{A}}_\lambda \,\ef{(n)} =: \mathcal{A}_{\lambda (n)}^{~~~(m)}\, \ef{(m)},~~~~~   \mathcal{A}_{\lambda (n)}^{~~~(m)}(q,p) \in C^\infty(\Gamma_{\te{s}},\mathbb{R})~~ \forall (n), (m) ,
\end{equation}
where the summation over $(m)$ includes essentially \emph{all} possible excitation numbers within the Fock space $\Hf$. However, there is only a countable number of $(m)$'s for which $\mathcal{A}_{\lambda (n)}^{~~~(m)}$ is non--vanishing if $(n)$ is a finite set of non--vanishing excitation numbers. For more details and the explicit calculations, we refer to \parencite{SchanderThiemannIV}. 

\subsection*{Application of a Space Adiabatic Perturbation Scheme}
The construction of the space adiabatic symbols is subject to two different perturbative scalings: $\pp$ for the homogeneous scalar field, and $\pp^2$ for the homogeneous gravitational degrees of freedom. The Moyal product for two operator--valued functions $\qf{f}(q,p) \in S^{m_1}_\rho(\Gamma_{\te{s}},\mathcal{B}(\Hf))$, $\qf{g}(q,p) \in S^{m_2}_\rho(\Gamma_{\te{s}},\mathcal{B}(\Hf))$ has the form,
\begin{equation}
(\qf{f}\star_\pp\qf{g})(q,p) \asymp \left[\qf{f} \exp \left( \frac{i\pp}{2} \left( \cev{\partial}_\phi \vec{\partial}_{p_\phi} - \cev{\partial}_{p_\phi} \vec{\partial}_\phi \right) - \frac{i\pp^2}{2} \left(\cev{\partial}_a \vec{\partial}_{p_a} - \cev{\partial}_{p_a} \vec{\partial}_a \right)\right) \qf{g} \right](q,p)~ \in S^{m_1+m_2}_\rho(\Gamma_{\te{s}},\mathcal{B}(\Hf))
\end{equation}
As it turns out, the Moyal product with respect to the gravitational degrees of freedom does not contribute to the computations up to second order in the perturbation scheme.

As before, the discrete eigenstate $\xi_{(\nu)_\beta}(q,p) \in \Hf$ with quantum number $(\nu)_\beta$ serves to define the zeroth order Moyal projector symbol, 
\begin{equation}
\qf{\pi}_{0}(q,p) := \sum_\beta \xi_{(\nu)_\beta}(q,p)\,\langle \xi_{(\nu)_\beta}(q,p),\cdot \,\rangle_{\te{f}}.
\end{equation}
The only relevant contribution to $\qf{\pi}_{1}$ comes from (S1--3). This off--diagonal part $\qf{\pi}_{1}^{\te{OD}}$ mixes the adjacent inhomogeneous eigenstates according to,
\begin{align}
\qf{\pi}_{1}^{\te{OD}} =&~ \frac{i}{2} \sum_{\beta =1}^\delta \sum_{(n)\neq (\nu)_\beta} \frac{A_{(\nu)_\beta (n)}}{E_{(\nu)_\beta} - E_{(n)}} \left( \xi_{(\nu)_\beta} \, \langle \xi_{(n)} , \cdot \rangle_{\te{f}} - \xi_{(n)} \, \langle \xi_{(\nu)_\beta} , \cdot \rangle_{\te{f}} \right), \\
\text{with}~~ A_{(\nu)_\beta (n)} =&~ \mathcal{A}_{\phi (\nu)_\beta}^{~~~(n)} \frac{\partial (E_{(n)} + E_{(\nu)_\beta})}{\partial p_\phi} - \mathcal{A}_{p_\phi (\nu)_\beta}^{~~~(n)}\, \frac{\partial (E_{(n)} + E_{(\nu)_\beta})}{\partial \phi} \nonumber \\
&~ + (E_{(n)}- E_{(m)}) \left(\mathcal{A}_{p_\phi (\nu)_\beta}^{~~~(m)} \,\mathcal{A}_{\phi (m)}^{~~~(n)} -\mathcal{A}_{\phi (\nu)_\beta}^{~~~(m)} \,\mathcal{A}_{p_\phi (m)}^{~~~~(n)}  \right).
\end{align}
Constructing the unitary symbol $\qf{\pi}_{(1)}$ requires to choose a simple reference space $\mathcal{K}_{\te{f}}$, and as before, $\Hf$ itself is a convenient choice. Its basis is determined by fixing a set of numbers $(q_0, p_0) \in \Gamma_{\te{s}}$ giving $\lbrace \zeta_{(n)} := \xi_{(n)}(q_0,p_0) \rbrace_{(n)}$. The zeroth order contribution to the Moyal unitary and the reference projector can be chosen as,
\begin{equation}
\qf{u}_{0}(q,p) := \sum_{(n)} \zeta_{(n)}\,\langle \xi_{(n)}(q,p),\cdot\,\rangle_{\te{f}},~~~~~
\qf{\pi}_{\te{R}} := \sum_{\beta =1}^\delta \zeta_{(\nu)_\beta} \langle \zeta_{(\nu)_\beta}, \cdot\, \rangle_{\te{f}}.
\end{equation}
The hermitian part of $\qf{u}_{1}(q,p)$ is determined by evaluating (S2--1) and (S2--2) up to first order in the perturbations,
\begin{equation}
\qf{u}_{1}^{\te{h}}(q,p) = \sum_{(n),(m),(k)} \left( \mathcal{A}_{\phi (n)}^{~~~(m)} \mathcal{A}_{p_\phi (m)}^{~~~~(k)} - \mathcal{A}_{p_\phi (n)}^{~~~(m)} \mathcal{A}_{\phi (m)}^{~~~~(k)} \right) \zeta_{(n)}\,\langle \zeta_{(k)}, \cdot \,\rangle_{\te{f}}.
\end{equation}
Since the sum runs over all possible combinations of quantum numbers, it is clear that the two contributions are equal and cancel each other. We thus have that $\qf{u}_{1}^{\te{h}}=0$. The antihermitian part of $\qf{u}_{1}$ results from employing the result for $\qf{\pi}_{1}^{\te{OD}}$ in the well--known expression from the toy model example,
\begin{equation}
\qf{u}_{1}^{\te{ah}} = \left[ \qf{\pi}_{\te{R}}, \qf{u}_{0}\cdot\qf{\pi}_{1}^{\te{OD}}\cdot \qf{u}_{0}^\ast \right]_{\te{f}} \cdot \qf{u}_{0}.
\end{equation}
We evaluate the effective Hamilton constraint symbol according to, $\qf{C}_{\te{eff}} = \qf{u} \star_\pp \qf{C} \star_\pp \qf{u}^\ast$, and restrict our interest directly to the reference space, i.e., to $\qf{C}_{\te{eff},\te{R}} = \qf{\pi}_{\te{R}} \cdot \qf{C}_{\te{eff}} \cdot \qf{\pi}_{\te{R}}$. At zeroth order, this yields,
\begin{align}
\qf{C}_{\te{eff},0,\te{R}} = \sum_{b,b' = 1}^{d,d'} &\left[ E_{\te{hom}}(a,p_a,\phi,p_\phi) + \frac{1}{a} \sum_{\vec{k} \in \mathbb{k}} \nu_{\te{MS},\vec{k},b} \omega_{\te{MS},\vec{k}} + \frac{1}{6a} \sum_{\vec{K} \in \mathbb{K}} \nu_{\te{T},\vec{K},b'} \omega_{\te{T},\vec{K}}\right] \nonumber \\
&~~ \cdot \zeta_{(\nu)_\beta} \, \langle \zeta_{(\nu)_\beta},\cdot\,\rangle_{\te{f}}, \label{eq:C zero}
\end{align}
which includes the standard zeroth order Hamilton constraint for an FLRW Universe $E_{\te{hom}}(a,p_a,\phi,p_\phi)$, and the bare energy contributions from the relevant energy band $\xi_{(\nu)_a}$. These additional terms are finite since the quantum numbers $\lbrace \nu_{\te{MS},\vec{k},b}, \nu_{\te{T},\vec{k},b'} \rbrace$ are non--vanishing for only a finite number of wave vectors $\vec{k}$. Te first order contribution to the effective Hamiltonian vanishes identically within the subspace of interest. 

The second order effective Hamilton symbol includes several contributions but only one of them is of second order in the perturbative parameter, and hence relevant. The occurence of terms that actually enter at higher orders in $\pp$ stems from the fact that the perturbative Mukhanov--Sasaki and graviton contributions to $\qf{C}$ are by definition of second order in $\pp$. It was necessary to include them to make the space adiabatic scheme work at the technical level. We refer again to \parencite{SchanderThiemannIV} for more details and only state the final result,
\begin{equation} \label{eq:Perturbative Model Ceff}
\qf{C}_{\te{eff},2,\te{R}}(a,p_a,\phi,p_\phi) = - \sum_{b = 1}^d \sum_{\vec{k} \in \mathbb{k}} \frac{1}{(\vec{k}^2 + M_{\te{MS}}^2)^{5/2}} \left(\nu_{\te{MS},\vec{k},b}+\frac{1}{2} \right) \frac{9}{2} \frac{m^4 p_\phi^4}{a^3 p_a^2} \, \zeta_{(\nu)_b} \langle \zeta_{(\nu)_b}, \cdot \rangle_{\te{f}}.
\end{equation}
This second order effective Hamiltonian symbol together with the zeroth order contribution \eqref{eq:C zero}, provides after Weyl quantization a constraint operator for the homogeneous sector of quantum gravity which includes, most importantly, the backreaction from the inhomogeneous modes. A similar result was obtained for a quantum cosmological model with scalar field perturbations and a deparametrizing dust particle, \parencite{SchanderThiemannIII}. The next step of the scheme consequently consists in Weyl quantizing the full effective constraint symbol and in finding physical quantum states on the homogeneous Hilbert space that are annihilated by it. A thorough discussion of the above results will be given in the next and final section.

\section{Discussion and Outlook}
\label{sec:Discussion and Outlook}
This review provides an introduction to the backreaction problem in classical, semiclassical and quantum cosmology, as well as a detailed overview of the current state of research in the respective fields. We have particularly focused on approaches to the backreaction problem in (perturbative) quantum cosmology that are inspired by Born--Oppenheimer methods. The main part of this paper is dedicated to a program which uses SAPT as due to \textcite{PST03}, and which extends the latter scheme to quantum field theoretical models. Thereby, it is possible to compute the backreaction effects from the quantum cosmological perturbations on the homogeneous and isotropic quantum background, \parencite{SchanderThiemannIII,SchanderThiemannIV}. We have advocated this framework here as it represents an unambigious and straighforward formalism in order to incorporate the yet neglected backreaction effects in quantum cosmology in a perturbative and rigorous way.

The extension of the SAPT methods actually requires some care. The first issue is related to a violation of the Hilbert--Schmidt condition in QFT on CST. In fact, it is well--known from standard QFT that Klein--Gordon fields with different masses give rise to unitarily inequivalent representations of the field algebra, \parencite{Haag1992}. Since here, the effective masses of the Klein--Gordon and tensor fields depend on the homogeneous FLRW background, the theory prevents unitarily equivalent quantum field theories for different background configurations. This would evidently impede the quantization of the homogeneous sector. \textcite{SchanderThiemannI,SchanderThiemannIV} show that it is possible to circumvent these problems by considering transformations of the whole system that are canonical up to second order in the cosmological perturbations, \parencite{CMM15,MartinezOlmedo2016,SchanderThiemannIII}. 

Another point is that the mass squared functions of the perturbative quantum fields that we denoted by $M_{\te{MS}}^2(a,p_a,\phi,p_\phi)$ and $(\pp M_{\te{T}})^2(a,p_a,\phi)$ in \eqref{eq:Cosmological Perturbations MS mass final} and \eqref{eq:Cosmological Perturbations Tensor mass final} are indefinite, and which leads to tachyonic instabilities. In \parencite{SchanderThiemannI}, several solutions are propsed, for example to revise the almost--canonical transformations that have actually led to these indefinite mass functions. A second proposal is to restrict the homogeneous phase space of the theory to regions in which the mass functions are positive. This can be made manifest by performing coordinate transformations in the slow sector. This is exemplified for the model with gauge--invariant perturbations in \parencite{SchanderThiemannI}. 

With the identification and solutions to these initial problems, it was possible to successfully apply the methods of SAPT to the backreaction problem in quantum cosmology. We stated the results for a cosmological, homogeneous and isotropic toy model, and for a fully--fledged perturbative quantum cosmology with gauge--invariant perturbations, \parencite{NeuserSchanderThiemannII,SchanderThiemannIV}.
In the first case, this effective Hamilton constraint includes the backreaction of the homogeneous scalar field; in the second case, the backreactions of the perturbative degrees of freedom on the homogeneous background are taken into account. Here, results up to second order in the adiabatic $\pp$--scheme are presented. The effective Hamiltonian symbol eventually needs to be quantized with respect to the slow sector and the goal is to find admissible solutions. This has been done for an oscillator toy model in \parencite{NeuserSchanderThiemannII}. One can proceed here in the same way but analytic solutions are harder to find due to the non--polynomial structure of the result, and which requires an in depth analysis of their dense domain. For simplicity, we chose a Weyl quantization scheme and a Schrödinger representation following the original work by \textcite{PST03}. Instead, one could consider the representation underlying LQC which could be of advantage regarding the domain issues, \parencite{Bojowald2008}. Due to the peculiarities of that representation (especially the strong discontinuity of the Weyl elements), and in agreement with certain superselection structures of the dynamics, one would need to discretize the labels of the Weyl elements in one of the conjugate variables. This would effectively replace the gravitational slow phase space $T^\ast\mathbb{R}$ by $T^\ast (S^1)$ for which the Weyl quantization in application to LQC has been discussed in \parencite{StottmeisterThiemann2016b}. 

Focussing on the second order contribution to the perturbative model, equation \eqref{eq:Perturbative Model Ceff}, one might be worried about the infinite sums. Note that the result splits into two parts, namely the one including the finite number of non--vanishing relevant quantum numbers $\nu_{\te{MS},\vec{k},b}$ for different degeneracy labels $b$, and the contributions which do not depend on these quantum numbers and hence include any summand of the wave vector sum. The first part has only a finite number of contributions and is manageable, while the second includes in principle an infinite sum. Fortunately, the wave vector square enters with an exponent of $-5/2$ which makes the sum a priori a convergent sum. But the effective Mukhanov--Sasaki mass squared $M_{\te{MS}}^2(a,p_a,\phi,p_\phi)$in the denominator is an \emph{indefinite} function on the homogeneous phase space. This will be cured as soon as a positive definite sector of the mass squared functions has been found. 

We emphasize again that the issue of convergence of the perturbation series in the SAPT approach has not been addressed here. We point to easily implementable strategies, \parencite{PST03,Stottmeister2015}, that allow to define auxiliary Hamiltonian symbols that capture the relevant physics of the model under consideration and whose perturbation series is safely convergent.

Finally, we stress that there is an obvious connection between backreaction and decoherence, \parencite{Schlosshauer2007}. Indeed, in decoherence, one aims at finding an effective description of what we call the slow sector using the reduced density matrix approach, tracing over the fast degrees of freedom, \parencite{Kiefer1987,PazSinha1991,PazSinha1992} (and references therein), and computing its effective dynamics, e.g., by solving associated Lindblad equations, \parencite{Manzano2020}. Using the tensor product structure of the full Hilbert space, the connection to our approach would be to construct the reduced density matrix from a density matrix on the full Hilbert space that can be formed from the eigenstates of the Hamiltonian (constraint) corresponding to a given energy band. Details will be given elsewhere.

\subsubsection*{Acknowledgments}
We thank Kasia Rejzner for helpful references and information on the backreaction problem within the algebraic approach to QFT. The work of S. Schander is supported by Perimeter Institute for Theoretical Physics. Research at Perimeter Institute is supported in part by the Government of Canada through the Department of Innovation, Science and Economic Development and by the Province of Ontario through the Ministry of Colleges and Universities.

\printbibliography

\end{document}